\documentclass[aps, twocolumn, superscriptaddress]{revtex4}

\makeatletter
\def\@dotsep{4.5}
\makeatother

\usepackage{graphicx}
\usepackage{amstext}
\usepackage{amsmath}
\usepackage{amssymb}
\usepackage{verbatim}
\usepackage{psfrag}
\usepackage{dcolumn}

\newcommand{\ud}{\mathrm{d}}
\newcommand{\bra}[1]{\langle #1|}
\newcommand{\ket}[1]{| #1\rangle}
\newcommand{\vect}[1]{\boldsymbol{#1}}
\newcommand{\eq}[1]{Eq.~\eqref{#1}}
\newcommand{\fig}[1]{Fig.~\ref{#1}}
\newcommand{\stn}[1]{Sec.~\ref{#1}}
\newcommand{\be}{\begin{equation}}
\newcommand{\ee}{\end{equation}}
\newcommand{\ti}[1]{\text{#1}}
\newcommand{\mc}[1]{\mathcal{#1}}
\newcommand{\w}{\omega}
\newcommand{\mean}[1]{\langle #1 \rangle}

\begin{document}

\title{Laser-induced currents along molecular wire junctions}
\author{Ignacio Franco}
\affiliation{Chemical Physics Theory
Group, Department of Chemistry, and  Center for Quantum Information and
Quantum Control, University of Toronto, Toronto, Ontario, Canada.}
\author{Moshe Shapiro}
\affiliation{Chemical Physics Department, The Weizmann Institute,
Rehovot, Israel, and Departments of Chemistry and Physics, The University of
British Columbia, Vancouver, B.C., Canada}
\author{Paul Brumer}
\affiliation{Chemical Physics Theory
Group, Department of Chemistry, and  Center for Quantum Information and
Quantum Control, University of Toronto, Toronto, Ontario, Canada.}

\date{\today}
\begin{abstract}
The treatment of the previous paper is extended to molecular 
wires. Specifically, the effect of electron-vibrational 
interactions on the electronic transport induced by femtosecond 
$\w+2\w$ laser fields along unbiased molecular nanojunctions is 
investigated.  For this,  the photoinduced vibronic dynamics  of  
\emph{trans}-polyacetylene oligomers coupled to macroscopic 
metallic leads is followed in a mean-field mixed 
quantum-classical approximation. A reduced description of the 
dynamics is obtained by introducing projective lead-molecule 
couplings and deriving an effective Schr\"odinger equation 
satisfied by the orbitals in the molecular region.  Two possible 
rectification mechanisms are identified and investigated. The 
first one relies on near-resonance photon-absorption and is shown 
to be fragile to the ultrafast electronic decoherence processes 
introduced by the wire's vibrations. The second one employs the 
dynamic Stark effect and  is demonstrated to be  highly efficient 
and robust to electron-vibrational interactions.
\end{abstract}

\maketitle
\section{Introduction}
Considerable  effort has been devoted to studies of   the  
properties of molecular wires~\cite{ lindsay_review, nitzan_review, 
ratner_review, adams_review, salomon_review, joachim_review, 
nitzan}, since they constitute natural candidates for the ongoing 
miniaturization of electronic devices.   Generally, the focus is 
placed  on the transport properties of  metal-molecule-metal 
junctions  subject to a bias voltage. In this regime  the  
metallic leads are the main source of electrons for  the 
transport while  the molecular system serves as a transporting 
medium that can be chemically functionalized to  modify the 
$I$-$V$ characteristics of the junction. 

Here we consider an alternative situation in which the junction  
is not subject to a bias voltage and where the molecule serves as 
the main source of transporting electrons. The composite system 
is taken to be spatially symmetric and net currents   are induced 
using  lasers with frequency components $\w$ and $2\w$. Such 
fields  give rise to phase-controllable transport in symmetric 
systems even when they have a zero temporal mean~\cite{paul, 
francoprl}. This rectification effect first appears in the third 
order response of the system to $\w+2\w$ fields. At this order 
the system mixes the frequencies and harmonics of the incident 
radiation generating a phase-controllable zero-frequency (DC) 
component in the response. The setup is of interest  since it can 
produce ultrafast currents and may lead to the development of 
molecular switches that operate on a femtosecond timescale. 

When using $\w+2\w$ fields to induce electronic transport  it is 
crucial to consider the influence of  the system's vibrations  on 
the laser scenario. This aspect becomes particularly relevant in 
molecular wires since  molecules nearly always exhibit vibronic 
couplings that can substantially modify  the photophysics of the 
junction and the effectiveness of the scenario.  This 
characteristic of molecular electronics marks an important 
difference relative to the behavior of rigid semiconducting 
solids, and is currently  the focus of vigorous  investigation 
~\cite{joachim_review, cizek, emberly, galperin_vibrations1, 
galperin_vibrations2, may, ness1, ness2, troisi}. 

In this paper we characterize the influence of vibronic couplings 
on  rectification   induced by femtosecond $\w+2\w$ laser  
pulses  along unbiased molecular wires. As a system we consider   
a \emph{trans}-polyacetylene (PA) oligomer  connected by its  
ends to macroscopic metallic leads. The leads are treated as 
rigid semi-infinite tight binding chains and the oligomer is 
described via  the  Su-Schrieffer-Heeger (SSH) 
Hamiltonian~\cite{HeegerAJ:NobLSm, SSH}. 

This work is a continuation of our studies  in 
Ref.~\onlinecite{papaper} in which we considered the problem of 
laser-inducing rectification along isolated PA chains. Naturally, 
it shares  many of the challenges encountered therein. In 
particular,  the problem of how to  effectively exert laser 
control on the electronic dynamics in the presence of 
lattice-induced  decoherence persists.  However, it  differs 
significantly from the case of isolated chains since  the 
coupling to the metallic contacts introduces a decay route for 
energetic electrons  and 
adds an additional element of complexity to the theoretical 
description.

Besides offering insight into vibrational effects in molecular 
electronics, this study contributes to the emerging direction of 
using lasers to modify the transport properties of nanojunctions 
in which several promising applications are being  recognized. 
For instance,  it has been demonstrated that light can  induce a 
switching behavior in the conduction  when, upon photoexcitation, 
the  molecule undergoes substantial conformational 
change~\cite{switch1, switch2, switchtheory}. In addition, recent 
theoretical work  suggests   that lasers may 
enhance~\cite{tikhonov_1, tikhonov_2} or 
suppress~\cite{hanggi_1vs2,  welack} transport induced by a bias 
or, alternatively, generate currents  in unbiased but asymmetric 
junctions~\cite{galperin_05, galperin_06, hanggi_1vs2}. 

Additional motivation for this study was provided by a rather 
puzzling observation made in Ref.~\cite{lehmann_1vs2}. In that 
work, the authors  analyzed vibrational effects on  laser-induced 
rectification along nanojunctions. The model presented there 
refers to a two-site molecular wire driven by a continuous wave 
$\w+2\w$ field. In it, each site is coupled perturbatively and 
independently to one local bath of harmonic oscillators in 
thermal equilibrium, and  an Ohmic spectral density is adopted  
to describe the phonon-induced relaxation.  This   work concludes 
that  vibronic couplings  \emph{enhance} the current induced by 
the symmetry breaking field by more than one order of magnitude. 
This observation seems to be in contradiction with our results in 
Ref.~\onlinecite{papaper} in which the vibronic couplings were 
shown to have strong detrimental effects on the control, 
specially when long pulses are employed. Further,   no insight 
into this rather nonintuitive behavior was provided.

In addition to clarifying this observation, and faithfully 
characterizing vibrational effects on laser rectification along 
molecular nanojunctions,  a novel mechanism to induce ultrafast 
currents along molecular wires that relies on the dynamic Stark 
effect instead of near-resonance photon absorption is discussed 
here~\cite{prlwire}. As shown below, this   mechanism 
is  remarkably robust to electron-vibrational couplings and is 
able to induce large currents in molecular wires with 
efficiencies $>90\%$. 

As in Ref.~\onlinecite{papaper}, the photoinduced vibronic 
dynamics of the wire is followed  in a mean-field mixed 
quantum-classical approximation~\cite{tully, halcomb, bornemann, 
prezdho}. Electronic decoherence and relaxation  is incorporated 
by averaging over an ensemble of quantum-classical trajectories 
with initial conditions obtained  by importance sampling the 
ground-state nuclear Wigner phase space distribution. In this 
manner, the simulations take into account the fact that the 
vibrational modes in typical molecular wires are nonlocal and 
constantly kept out of thermal equilibrium either through direct 
interaction with the field or through exchange of energy with the 
driven electrons. 

A reduced description of the dynamics of the composite system is 
obtained by deriving an effective Schr\"odinger equation 
satisfied by the orbitals in the molecular region. As shown, in 
the wide bandwidth limit the lead regions are mapped into 
negative imaginary (absorbing) potentials. In addition,  we 
employ projective lead-molecule couplings  that incorporate  
basic aspects of the Fermi sea while avoiding the necessity to 
explicitly follow the dynamics of the essentially infinite number 
of electrons in the leads.   This model is only appropriate to 
describe the short time dynamics of unbiased junctions since, in 
it, charge is not permitted to flow from the leads into the 
oligomer.


This manuscript is organized as follows: In \stn{wiresec:methodology} we briefly describe the model (\stn{wiresec:model}), and  determine the effective mean-field equations of motion for the nanojunction (\stn{wiresec:eom}).  Subsequently, we derive an expression for the photoinduced currents (\stn{wiresec:obs}) that solely depends on molecular properties. Our main results are stated in \stn{wiresec:results}. Specifically, in \stn{wiresec:photonabsorption} we study the scenario in its  usual regime where the laser frequencies are tuned at or near resonance with one of the electronic transition frequencies. Subsequently, in \stn{wiresec:stark}, we introduce the Stark shift rectification mechanism. The  emerging trends are  summarized  in \stn{wiresec:conclusions}.

\section{Model and methods}
\label{wiresec:methodology}
\subsection{Model}
\label{wiresec:model}
As a system we consider a \emph{trans}-polyacetylene oligomer 
connected by its ends to macroscopic metallic leads.  Since 
typical metals screen electric fields that have a frequency below 
the plasma frequency, so that  electromagnetic radiation from the 
optical or the infrared spectral range is almost perfectly 
reflected at the surface~\cite{fox}, we  assume that the 
radiation field only interacts with the oligomer chain.
 
 The 
composite system is described as  a one-dimensional lattice  in 
which each site $n$  corresponds to the position of an atom, and 
is defined  by the  Hamiltonian 
\be
\label{wireeq:tothamcomposite}
H(t) = H_\ti{L} + H_\ti{S-L} + H_\ti{S}(t) + H_\ti{S-R} + 
H_\ti{R}. 
\ee The left (L) and right (R) leads are treated as 
semi-infinite rigid tight-binding chains  with hopping parameter 
$t_\ti{lead}$, so that 
\be \label{wireeq:hamleads}
\begin{split}
H_\ti{L} &= -t_\ti{lead}\sum_{n<0} \sum_{s=\pm 1}
(c_{n+1,s}^\dagger c_{n,s} + c_{n,s}^\dagger c_{n+1,s}), \\
H_\ti{R} & = -t_\ti{lead}\sum_{n>N} \sum_{s=\pm 1}
(c_{n+1,s}^\dagger c_{n,s} + c_{n,s}^\dagger c_{n+1,s}).
\end{split} 
\ee
 Here $c_{n,s}^{\dagger}$ ($c_{n,s}$)  creates (annihilates) a 
fermion in site $n$ with spin $s$ and satisfies  the usual 
fermionic anticommutation relations 
$\{c_{n,s},c_{m,s'}^{\dagger}\}=\delta_{n,m}\delta_{s,s'}$.  The 
$N$-membered oligomer chain  with Hamiltonian $H_\ti{S}(t)= 
H^\ti{el}_\ti{S}(t)  + H_\ti{S}^\ti{ph}(t)$  is situated between 
sites  $n=1, \ldots, N$ and is described by the SSH 
model~\cite{HeegerAJ:NobLSm, SSH}  coupled to a laser field 
$E(t)$ in dipole approximation.  Specifically, the electronic 
part of $H_\ti{S}$ is 
\be
\begin{split}
\label{wireeq:elec}
H_\ti{S}^\ti{el}(t) = \sum_{n=1,s}^{N-1} [-t_0+\alpha(u_{n+1}-u_n)]
(c_{n+1,s}^\dagger c_{n,s} \\
+ c_{n,s}^\dagger c_{n+1,s}) 
+ |e| \sum_{n=1,s}^{N} x_{n} c_{n,s}^\dagger
c_{n,s} E(t),
\end{split}
\ee
 where  $t_0$ is the hopping integral for zero displacement, $u_n$ is the
monomer displacement of site $n$ from its perfectly periodic position $na$, and $\alpha$ describes the electron-ion
coupling between neighboring sites. In turn, $x_n=(na+ u_n)$ is the position
operator for site $n$,  $a$ the lattice constant, and  $-|e|$  the unit
electronic charge.   The lattice is described by
\begin{equation}
H_\ti{S}^{\ti{ph}}(t) = \sum_{n=1}^{N} \frac{p_n^2}{2M} +
\frac{K}{2}\sum_{n=1}^{N-1} \left(u_{n+1} - u_{n}\right)^{2}
-|e|\sum_{n=1}^{N} x_n E(t),
\end{equation}
with force constant $K$ and (CH) group mass $M$, and where $p_n$ 
is the momentum conjugate to $u_n$.
 Last, the coupling between the oligomer and the leads
\be
\label{wireeq:lead-molecule}
\begin{split}
H_{\ti{S-L}} & = -t_\ti{coup} \sum_{s=\pm 1} (c_{1,s}^\dagger c_{0,s} +
 c_{0,s}^\dagger c_{1,s} )\\
H_{\ti{S-R}} & = -t_\ti{coup} \sum_{s=\pm 1} (c_{N+1,s}^\dagger c_{N,s} + c_{N,s}^\dagger c_{N+1,s} ),
\end{split}
\ee is taken to be at a single site only with coupling constant 
$t_\ti{coup}$.

\subsection{Equations of motion}
\label{wiresec:eom}
The electron-vibrational dynamics of the composite system in the 
presence of a radiation field is followed in the mean-field (Ehrenfest) mixed 
quantum-classical approximation introduced in Sec. IIB of Ref.~\onlinecite{papaper}. As before, the nuclei satisfy classical trajectories governed by
\begin{equation}
\label{wireeq:nuclei}
\begin{split}
\dot{u}_n(t)  = &\frac{p_n(t)}{M}; \\
\dot{p}_n(t)  = & - K\left(2u_n(t) - u_{n+1}(t) - u_{n-1}(t)\right)  \\ 
&+ 2\alpha\textrm{Re}\left\{ \rho_{n, n+1}(t)
-   \rho_{n, n-1}(t) \right\}\\
& - |e|E(t) \left( \rho_{n,n}(t) -1\right),
\end{split}
\end{equation}
where $\rho_{n,m}(t) = \sum_s \bra{\Psi(t)} c_{n,s}^\dagger c_{m,s} \ket{\Psi(t)}$ is the reduced electronic density matrix. The chain is taken to be clamped so that $u_1(t)=u_N(t)=0$ and $p_1(t)=p_N(t)=0$, and \eq{wireeq:nuclei} is valid for $n=2, \cdots, N-1$.  In turn, the many electron wavefunction $\ket{\Psi(t)}$ satisfies the Schr\"odinger equation and, hence, the dynamics of $\rho_{n,m}(t)$ is given by
\be
\label{wireeq:rdmdyn}
i\hbar \frac{\ud}{\ud t} \rho_{n,m}(t) = \sum_{s} \bra{\Psi(t)} 
[c_{n,s}^\dagger c_{m,s}, H_\ti{elec}(t)] \ket{\Psi(t)},
\ee 
where 
\be
\label{eq:totelec}
H_\ti{elec}(t) = H_\ti{L} + H_\ti{S-L} + H_\ti{S}^\ti{el}(t) + 
H_\ti{S-R} + H_\ti{R} 
\ee 
is the electronic part of the Hamiltonian.  For 
an initial electronic state  $\ket{\Psi(0)}$  that is well 
described by a single Slater determinant, the reduced density 
matrix admits the orbital decomposition 
\be
\label{wireeq:rdm}
\rho_{n,m}(t) = \sum_{\epsilon, s}  \bra{\epsilon(t),s}n,s\rangle \langle m,s\ket{\epsilon(t),s}f(\epsilon,s)
\ee
where $f(\epsilon, s)$ is the time-independent initial 
distribution function that takes values 0 or 1 depending on the 
occupation of each level with energy $\epsilon$ and spin $s$, and 
$\ket{n,s} = c_{n,s}^\dagger\ket{0}$ where $\ket{0}$ represents 
the vacuum state.  In order for $\rho_{n,m}(t)$ to satisfy 
\eq{wireeq:rdmdyn}, the  orbitals $\ket{\epsilon(t),s}$ must be 
solutions to the single-particle time-dependent Schr\"odinger 
equation \be
\label{wireeq:schrodinger}
i\hbar \frac{\ud}{\ud t} \ket{\epsilon(t), s} = H_\ti{elec}(t) 
\ket{\epsilon(t),s} \ee with initial condition 
$\ket{\epsilon(t=0),s} = \ket{\epsilon,s}$, where 
$\ket{\epsilon,s}$ denote the eigenorbitals of $H_\ti{elec}$ at 
preparation time.

In principle, by integrating  the above equations self-consistently  it is possible to 
follow the coupled dynamics of both electronic and nuclear 
degrees of freedom in the composite system. In practice, the 
problem is intractable since the number of equations to be 
integrated is unbounded due to the 
large number of electrons in the leads and the spatial 
unboundedness of the composite system.  Further, the simplifications that have 
been developed for  periodic systems are not applicable here. 
 A reduced description 
of the dynamics is obtained by  first deriving, in 
\stn{sec:nonmarkovianse}, an  effective single-particle 
Schr\"odinger equation  that is satisfied   in the molecular 
region. This eliminates the necessity of integrating 
\eq{wireeq:schrodinger} over the complete spatial domain of the 
composite system.   Subsequently, in \stn{stn:projective}, we 
specialize the coupling between the molecule and the leads using 
projection operators so that basic aspects of the lead-molecule 
coupling are captured without explicitly following the dynamics 
of the electrons originally in the leads. Then, in 
\stn{wiresec:umatrixelements},  we derive an expression for the 
memory kernel of the resulting equations, which  is then employed 
in \stn{stn:bandwidth} to obtain the  Markovian limit  of the 
dynamics.

\subsubsection{Effective non-Markovian Schr\"odinger equation}
\label{sec:nonmarkovianse}
Henceforth, we drop spin labels since orbitals with the same 
energy but opposite spin satisfy the same equation of motion 
under $H_\ti{elec}(t)$. We begin by partitioning  the orbitals 
that form $\rho_{n,m}(t)$ into the three spatial regions of the 
composite system \be
\label{wireeq:partition}
\ket{\epsilon(t)} = \ket{\epsilon_\ti{L}(t)} + 
\ket{\epsilon_\ti{S}(t)} +  \ket{\epsilon_\ti{R}(t)}. \ee 
Here 
$\ket{\epsilon_{\ti{P}}(t)} =  \sum_{n\in \ti{P}} 
\bra{n}\epsilon(t)\rangle \ket{n}$ describes the part of the 
orbital in region P, where P can be   the left lead (L), right 
lead (R) or system (S).  In terms of this partition  the 
single-particle Schrodinger equation~\eqref{wireeq:schrodinger} 
reads
\be
\label{wireeq:seorbital}
i\hbar \frac{\ud}{\ud t}
\begin{bmatrix}
\ket{\epsilon_{\ti{L}}(t)} \\
\ket{\epsilon_\ti{S}(t)} \\
\ket{\epsilon_\ti{R}(t)}
\end{bmatrix}
=
\begin{bmatrix}
H_{\ti{L}} & H_\ti{S-L}(t) & 0 \\
H_\ti{S-L}(t) & H_\ti{S}^{\ti{el}}(t) & H_\ti{S-R}(t) \\
0 & H_\ti{S-R}(t) & H_\ti{R}
\end{bmatrix}
\begin{bmatrix}
\ket{\epsilon_{\ti{L}}(t)} \\
\ket{\epsilon_\ti{S}(t)} \\
\ket{\epsilon_\ti{R}(t)}
\end{bmatrix},
\ee where we have explicitly allowed for time-dependence in the 
lead-molecule coupling since this characteristic will arise when 
introducing projection operators into the dynamics.

We seek  an effective equation of motion for  
$\ket{\epsilon_\ti{S}}$ in which it is not necessary to 
explicitly follow the dynamics of $\ket{\epsilon_\ti{L}}$ or 
$\ket{\epsilon_\ti{R}}$~\cite{frensley}.  This can be 
accomplished by Laplace  transforming the evolution equation for 
the lead components,  solving the resulting algebraic equations 
and  inverse Laplace transforming by taking into account the 
convolution theorem. The procedure yields:
\begin{equation}
\label{wireeq:leadorbital}
\ket{\epsilon_\beta(t)}
=  U^\beta(t) \ket{\epsilon_\beta(0)}
+ \frac{1}{i\hbar} \int_0^t  U^\beta(t-\tau)
H_{\ti{S-}\beta}(\tau) \ket{\epsilon_\ti{S}(\tau)}\ud\tau,
\end{equation}
where $\beta=$ L or R denotes the left or right lead, 
respectively.   The first term in \eq{wireeq:leadorbital}  
describes the evolution of the lead components of the orbitals in 
the absence of coupling to the molecular system. Naturally, the 
dynamics is determined by  the evolution operator for the 
isolated  leads $U^\beta (t) =  \exp({-i H_{\beta} t/\hbar})$.   
In turn, the influence of the lead-molecule coupling is mapped 
into a convolution integral with $U^\beta(t-\tau)$  acting as the 
memory kernel. 

By introducing  \eq{wireeq:leadorbital} into the evolution 
equation for $\ket{\epsilon_\ti{S}}$ we arrive at  an effective 
non-Markovian Schr\"odinger equation for the system part of the 
orbital: \be
\begin{split}
\label{wireeq:nonmarkovianse}
i\hbar \frac{\ud}{\ud t} \ket{\epsilon_\ti{S}(t)}   =   & H_\ti{S}(t)\ket{\epsilon_\ti{S}(t)}
+    \sum_{\beta=\ti{L, R}}\Big[
 H_{\ti{S-}\beta}(t) U^{\beta}(t) \ket{\epsilon_{\beta}(0)} \\
 & +  \frac{H_{\ti{S-}\beta}(t)}{i\hbar} \int_{0}^{t}
U^{\beta}(t-\tau) H_{\ti{S-}\beta}(\tau)
\ket{\epsilon_\ti{S}(\tau)}\ud \tau
 \Big].
\end{split}
\ee The first term corresponds to  the single-particle 
Schr\"odinger equation for the isolated system. The remaining 
terms  describe the transfer of population amplitude between the 
system and the leads. 

The advantage of \eq{wireeq:nonmarkovianse} over 
\eq{wireeq:schrodinger} is that it permits following the dynamics 
of the composite system by integrating the equations of motion in 
a finite spatial region.  The influence of the leads  has been   
mapped into a term that is nonlocal in time and a residual term 
dependent on the initial state of the leads.  
Equation~\eqref{wireeq:nonmarkovianse} is also a good starting 
point for various approximations. In particular, the Markovian 
limit arises naturally when the energy bandwidth of the metallic 
leads is much larger than any other characteristic energy of the 
system.

\subsubsection{Projective molecule-lead coupling}
\label{stn:projective}
Solving the above integro-differential equation 
[\eq{wireeq:nonmarkovianse}] exactly  for all the initially 
occupied levels in the composite system is a problem as 
formidable as the original one, and an  effective way to 
introduce the  leads into the dynamics is required.  Physically,  
the  contacts  block any electrons from the molecule  with energy 
less than the Fermi energy and act as a thermal source of 
electrons to the molecular system. We are interested in the  case 
where there is no bias voltage across the bridge and in short 
simulation times that limit the degree of relaxation of electrons 
from the leads to the   molecular states. In this regime, the 
main effect of the leads is to accept electrons  that have energy 
higher than the Fermi energy $\epsilon_\ti{F}$ and to block them 
otherwise. We model this property using  projection operators 
that restrict the interaction between the molecule and the leads 
to only those molecular levels that are above the Fermi energy.

Consider a projection operator ($\mc{P}^2 = \mc{P}$) \be 
\mc{P}(t) = \mc{P}^\ti{S}(t) + \mc{P}^\ti{L} +\mc{P}^\ti{R} \ee
 where
\be
 \mc{P}^\beta = \sum_{\epsilon\in \beta}
\ket{\epsilon_\beta}\bra{\epsilon_\beta}, \ee with $\beta=$L or 
R, projects onto single-particle states of the  left/right lead. 
In turn, \be
\label{wireeq:sysprojection}
\mc{P}^\ti{S}(t) = \sum_{\epsilon_\gamma > \epsilon_\text{F}} 
\ket{\gamma_\ti{S}(t)}\bra{\gamma_\ti{S}(t)} \ee is a projection 
operator onto  the instantaneous light-dressed eigenorbitals 
$\ket{\gamma_\ti{S}(t)}$ of the molecular system that reside 
above the Fermi energy. The latter are  defined by the eigenvalue 
relation \be
 H_{\ti{S}}^\ti{el}(t) \ket{\gamma_\ti{S}(t)}   = \epsilon_{\gamma}(t) \ket{\gamma_\ti{S}(t)},
\ee where $H_{\ti{S}}^\ti{el}(t)$ is the  electronic Hamiltonian 
of the oligomer  at time $t$  given in \eq{wireeq:elec}. The 
time-dependence  of  $\ket{\gamma_\ti{S}(t)}$  [and  
$\mc{P}^\ti{S}(t)$] is a result of the time-dependence in the 
electronic Hamiltonian introduced by the field and the nuclear 
motion.  An evident but important property of the projection 
operator is  that $\mc{P}^\ti{S} \mc{P}^\beta=\mc{P}^\beta 
\mc{P}^\ti{S}=0$. 

We now wish to use $\mc{P}$ to restrict the lead-molecule 
interaction. The projection operator cannot be  introduced 
directly into  the second quantized version of the  lead-molecule 
coupling [\eq{wireeq:lead-molecule}]. Rather, it is necessary to 
modify the first-quantized interaction Hamiltonian and then 
second quantize it.  In first quantization, the original coupling 
takes the form $H_{\ti{S-}\beta}= \sum_{i=1}^{\mc{N}} 
h_{\ti{S-}\beta}(i)$, where the label $i$ refers to the 
coordinates of the $i$-th electron and  the sum runs over all 
$\mc{N}$ electrons in the system. The nearest-neighbor couplings 
of \eq{wireeq:lead-molecule}, give rise to the following matrix 
elements of $h_{\ti{S-}\beta}$, \be
\label{wireeq:couplingmatrixelements}
\begin{split}
\bra{n} h_\ti{S-L} \ket{m} & = -t_\ti{coup} (  \delta_{n,1}\delta_{m,0} + \delta_{n,0}\delta_{m,1} ), \\
\bra{n} h_\ti{S-R} \ket{m} & = -t_\ti{coup} ( \delta_{n,N}\delta_{m,N+1} + \delta_{n,N+1}\delta_{m,N}). \\
\end{split}
\ee By introducing $\mc{P}$ into the first-quantized coupling 
Hamiltonian, so that \be H_{\ti{S-}\beta}(t) = 
\sum_{i=1}^\mc{N}\mc{P}^\dagger(i,t) h_{\ti{S-}\beta} (i) 
\mc{P}(i,t), \ee only those electrons in the desired  subset of 
electronic levels are allowed  to move from the molecule to the 
leads, and vice versa. The modified coupling operator is 
Hermitian, and thus diagonalizable, and hence can be cast into 
second quantized form using the usual procedure~\cite{negele}, 
to  obtain \be
\begin{split}
H_\ti{S-L}(t)  & = -t_\ti{coup} \sum_{n,m,s}\mc{P}_{n,0}(t)\mc{P}_{1,m}(t) c_{n,s}^\dagger c_{m,s} + \text{H.c}, \\
H_\ti{S-R}(t)  & = -t_\ti{coup} \sum_{n,m, s} \mc{P}_{n,N}(t)\mc{P}_{N+1,m}(t)  c_{n,s}^\dagger c_{m,s} + \text{H.c.},
\end{split}
\ee where $\mc{P}_{n,m} = \bra{n}\mc{P}\ket{m}$ and  H.c. stands 
for Hermitian conjugate.

We specialize our considerations to the case in which electrons 
in the molecule are allowed to leak into the leads but where no 
electrons that are originally in the leads can   enter the 
molecule. This is  modeled by treating the leads as electron-less 
and  by coupling the excited electrons in the molecular system to 
every level in the leads. Since we now project onto a complete 
lead basis $\mc{P}_{n,0}  = \delta_{n,0}$ and $\mc{P}_{n,N+1}  = 
\delta_{n,N+1}$, and the  lead-molecule interaction terms assume 
the  simplified form: \be
\label{wireeq:projectedcoupling}
\begin{split}
H_\ti{S-L}(t)    & = -t_\ti{coup} \sum_{n\in \ti{S}}\sum_{s=\pm 1} \mc{P}^\ti{S}_{1,n}(t) c_{0,s}^\dagger c_{n,s} +  \text{H.c.}, \\
H_\ti{S-R}(t)     &= -t_\ti{coup} \sum_{n\in \ti{S}} \sum_{s=\pm1} \mc{P}^\ti{S}_{N,n}(t) c_{N+1,s}^\dagger c_{n,s} + \text{H.c.}. \\
\end{split}
\ee 

The leads and molecule are taken to be initially detached and, 
since the projective coupling \eq{wireeq:projectedcoupling} 
already incorporates the Fermi blockade imposed by the leads, we 
only consider orbitals that are initially in the molecular region 
so that  $\ket{\epsilon_\beta(t=0)}=0$   for the states of 
interest. Taking into account the above, and introducing the  
projective coupling \eqref{wireeq:projectedcoupling} into 
\eq{wireeq:nonmarkovianse},  it  follows that
\begin{multline}
\label{wireeq:nmseprojected}
i\hbar \frac{\ud}{\ud t} \langle n \ket{\epsilon_\ti{S}(t)}  =
\sum_{m=1}^{N}\bra{n} H_\ti{S}^\ti{el}(t)\ket{m} \langle m \ket{\epsilon_\ti{S}(t)} \\
+ \frac{t_\ti{coup}^2}{i\hbar}\sum_{m=1}^{N} \int_{0}^{t}  \,
\big[\mc{P}_{n,1}^{\ti{S}}(t) U^\ti{L}_{0,0}(t-\tau)\mc{P}_{1,m}^{\ti{S}}(\tau)\\
 +
\mc{P}_{n,N}^{\ti{S}}(t) U^\ti{R}_{N+1,N+1}(t-\tau)\mc{P}_{N,m}^{\ti{S}}(\tau)
 \big]  \langle m\ket{\epsilon_\ti{S}(\tau)} \ud \tau.
\end{multline}

\subsubsection{The memory kernel}
\label{wiresec:umatrixelements}
To proceed further  an explicit expression for the memory kernels 
in \eq{wireeq:nmseprojected} is required. The task is to find 
the  matrix elements of the evolution operator 
$U^\beta(t)=\exp(-i H_\beta t/\hbar)$  for the left ($\beta=$L) 
and right ($\beta=$R) semi-infinite tight binding chains in site 
representation.  For this, first note that the lead Hamiltonian 
[\eq{wireeq:hamleads}] can be diagonalized by performing the 
basis transformation: \be
\label{wireeq:basistransformation}
c_{n,s} = \left(\frac{2}{\Omega}\right)^{1/2} \sum_{k=1}^{\Omega} 
\sin\left[(n-n_{\beta})k \pi/\Omega\right] c_{k,s}, \ee where 
$\Omega$ is the number of sites in the leads and $n_{\beta}$ is 
the molecular site that is connected to  lead $\beta$, i.e.  
$n_\ti{L}=1$  and $n_\ti{R}=N$.   
Inserting~\eq{wireeq:basistransformation} into 
\eq{wireeq:hamleads} yields \be
\label{wireeq:leaddiagonalized}
H_{\beta} = -2t_\ti{lead} \sum_{k=1}^{\Omega} \sum_{s=\pm 1} 
\cos\left({k \pi}/{\Omega}\right) c_{k,s}^\dagger c_{k,s} \ee 
where we have exploited the fact that for large $\Omega$, 
$\delta_{k,k'}=\frac{1}{\Omega} \sum_{n=1}^{\Omega-1} \exp{(i 
n(k-k')\pi/\Omega)}$. 

The matrix elements of the evolution operator in site 
representation $n$ are then given by:
\begin{equation*}
\begin{split}
U_{n,m}^\beta(t)  & =
\bra{n} U^\beta(t)\ket{m} \\
 & =  \frac{2}{\Omega} \sum_{k=1}^{\Omega}
\sin\left[{(n-n_\beta) k \pi}/{\Omega}\right] \times \\
& \sin\left[{(m-n_\beta) k \pi}/{\Omega}\right]
   \exp{\left[2 i  t_\ti{lead}
        \cos{\left({k\pi}/{\Omega} \right)}t/\hbar\right]},
\end{split}
\end{equation*}
where we have used Eqs. \eqref{wireeq:basistransformation} 
and~\eqref{wireeq:leaddiagonalized}. For  large $\Omega$ the sum 
can be approximated by an integral to give: 
\be
\label{wireeq:matrixelements}
\begin{split}
U_{n,m}^\beta(t)  = & - \frac{1}{\pi}\int_0^\pi  \big\{
\cos \left[(n+m-2n_\beta)\theta\right] \\
& - \cos\left[ (n-m)\theta\right]  \big\}
\exp{\left[2 i t_\ti{lead} \cos(\theta) t/\hbar  \right]}\, \ud \theta \\
= & \,i^{n-m}  J_{n-m}\left(\frac{2 t_\ti{lead}}{\hbar} t\right) \\
   & -  i^{n+m-2n_\beta}  J_{n+m-2n_\beta}\left(\frac{2 t_\ti{lead}}{\hbar} t\right),
\end{split}
\ee 
where $J_{n}(z)= \frac{i^{-n}}{\pi}\int_{0}^{\pi} e^{i z 
\cos{\theta}} \cos(n \theta)\, \ud\theta$ is a Bessel function of 
the first kind of order $n$~\cite{bessel:abramowitz}. We note 
that $U_{n,m}^\beta$  satisfies the equations of motion, as well 
as the boundary and initial conditions:
\begin{equation*}
\begin{split}
i\hbar \frac{\ud}{\ud t}U_{n,m}^\beta(t) & = \sum_{r\in\beta} \bra{n} H_{\beta} \ket{r} U_{r,m}^\beta(t); \\ 
U_{n_\beta, m}^\beta(t)   & = U_{m,  n_\beta}^\beta(t)=0;  \quad  U_{n, m}^\beta(t=0) = \delta_{n,m}.
\end{split}
\end{equation*}

Of particular relevance  is \be
\label{wireeq:kernelfinal}
\begin{split}
\mc{K}(t) \equiv U_{0,0}^\ti{L}(t)= U_{N+1,N+1}^\ti{R}(t)
& =  \frac{2 J_{1}\left(\frac{2 t_\ti{lead}}{\hbar} t\right)}{\frac{2 t_\ti{lead}}{\hbar} t},
\end{split}
\ee where we have used the identities $J_{-n}(z) = (-1)^n 
J_{n}(z)$ and  $J_{n-1}(z) + J_{n+1} (z) = \frac{2n}{z} J_n(z)$. 
This quantity determines the kernel of the convolution integral 
in \eq{wireeq:nmseprojected}.  It is 
characterized by a pronounced  peak around $t=0$ followed by  
oscillations that decay asymptotically as  
$\frac{1}{\sqrt{\pi}}\left(\frac{\hbar}{t_\ti{lead} 
t}\right)^{3/2}$. This asymptotic dependence is exploited in the 
next section to derive the Markovian limit of the equations of 
motion.

\subsubsection{Wide bandwidth approximation}
\label{stn:bandwidth}
We now invoke the wide bandwidth approximation in which the 
energy bandwidth of the leads ($4t_\ti{lead}$) becomes the 
dominant energy in the problem and, consequently, 
$\hbar/t_\ti{lead}$  the fastest time scale. In this regime, the 
memory kernel  $\mc{K}(t-\tau)$ becomes  sharply peaked around 
$\tau=t$ and the orbitals do not change appreciably on the time 
scale in which $\mc{K}(t-\tau)$ varies. It follows that the 
convolution integrals in \eq{wireeq:nmseprojected} can be 
approximated by
\begin{equation*}
\begin{split}
 \int_{0}^{t} \mc{P}_{n,k}^{\ti{S}}(t) \mc{K}(t-\tau)\mc{P}_{k,m}^{\ti{S}}(\tau)   \langle m \ket{\epsilon_\ti{S}(\tau)}\ud \tau \approx \\
  \mc{P}_{n,k}^{\ti{S}}(t)\mc{P}_{k,m}^{\ti{S}}(t) \langle m \ket{\epsilon_\ti{S}(t)}\int_{0}^{t} \mc{K}(t-\tau)   \ud \tau.
\end{split}
\end{equation*}
Further, for times $ t \gg \frac{\hbar}{t_\ti{lead}}$ the 
remaining time integral is well approximated by its asymptotic 
value,  $\lim_{t\to\infty} \int_{0}^{t} \mc{K}(t-\tau)   \ud \tau 
= \hbar/t_\ti{lead}$, and the equations of motion reduce to their 
Markovian limit: 
\be
\begin{split}
\label{wireeq:finalorbs}
i\hbar \frac{\ud}{\ud t}\langle n \ket{\epsilon_\ti{S}(t)} =
\sum_{m=1}^{N}\left[\bra{n} H_\ti{S}^\ti{el}(t)\ket{m}   - i 
 \frac{t_\ti{coup}^2}{t_\ti{lead}} \Gamma_{n,m}\right]\langle m 
\ket{\epsilon_\ti{S}(t)}, 
\end{split}
\ee where 
\be
\label{wireeq:gamma}
\Gamma_{n,m}(t) = 
\mc{P}_{n,1}^{\ti{S}}(t)\mc{P}_{1,m}^{\ti{S}}(t) 
+\mc{P}_{n,N}^{\ti{S}}(t) \mc{P}_{N,m}^{\ti{S}}(t). 
\ee 
Equation~\eqref{wireeq:finalorbs}  provides a realistic mechanism 
for electron absorption and  energy dissipation in the dynamics 
of a metal-molecule-metal system.  The influence of the leads in 
the dynamics has been mapped into a negative imaginary 
(absorbing) potential. The projection operators in the coupling, 
contained within $\Gamma_{n,m}$,  ensure that only those 
electrons with sufficient energy get absorbed, with proper 
conservation of the  antisymmetry principle of the many-electron 
wavefunction. The equation is valid for times $t\gg 
\hbar/t_\ti{lead}$ but can be applied  for all times by turning 
on  slowly the interaction between the molecule and the leads; a 
strategy that we adopt. 

Note that the  field influences the dynamics directly  by 
modifying the wire component, and  indirectly by influencing the 
wire-lead coupling since  $\Gamma_{n,m}(t)$ depends both on the 
nuclear dynamics  and the radiation-matter interaction. This 
characteristic introduced by the field has also been observed in  
an alternative description of the laser-induced dynamics of 
molecular wires~\cite{welack}. Further note that this  model 
neglects hole transport. Nevertheless, since the SSH Hamiltonian 
is electron-hole symmetric, so that a subsequent inclusion of 
this contribution is just expected to double the resulting 
current.

\subsubsection{Final equations}
Consider now the final equations of motion. The  nuclear degrees 
of freedom satisfy trajectories determined by \eq{wireeq:nuclei}. 
The orbitals determining  $\rho_{n,m}(t)$ [\eq{wireeq:rdm}] are 
the initially occupied molecular orbitals which, by virtue of 
\eq{wireeq:finalorbs}, satisfy
\begin{equation}
\label{wireeq:elec2}
\begin{split}
 i\hbar \frac{\ud}{\ud t} \langle n \ket{\epsilon_\ti{S}(t)}
& =  \left[-t_0 + \alpha (u_{n+1}(t)-u_n(t))\right]\langle n+1\ket{\epsilon_\ti{S}(t)}  \\
& +  \left[-t_0 + \alpha (u_{n}(t)-u_{n-1}(t))\right]\langle{ n-1}\ket{\epsilon_\ti{S}(t)} \\
& +   |e|E(t) \left(na+u_n(t)\right) \langle n\ket{\epsilon_\ti{S}(t)} \\
& - i \frac{t_\ti{coup}^2}{t_\ti{lead}}  \sum_{m=1}^{N} \Gamma_{n,m}(t)\langle n \ket{\epsilon_\ti{S}(t)},
\end{split}
\end{equation}
 for $n=2, \cdots, N-1$. At the boundaries of the oligomer, $n=1$ and $n=N$, the orbitals satisfy \eq{wireeq:elec2} without the term that involves the amplitude of the orbital amplitudes outside the molecular domain.   Equations~\eqref{wireeq:nuclei}  and~\eqref{wireeq:elec2}, together with the auxiliary quantity \eqref{wireeq:rdm}, constitute a closed set of  $N(N+2)$  coupled first-order differential equations that are integrated using a Runge-Kutta method  of order eight~\cite{rksuite}. The projective term in the dynamics  $\Gamma_{n,m}(t)$ is updated at every time step  by diagonalizing the system's electronic Hamiltonian $H_\ti{S}^\ti{el}(t)$  to obtain the instantaneous eigenstates $\ket{\gamma_\ti{S}(t)}$  defining the projection operator  through \eq{wireeq:sysprojection}. 
We project onto states  that are above the Fermi energy, taken to be the zero  reference energy. In other words, out of the $N$ single-particle instantaneous eigenstates of  $H_\ti{S}^\ti{elec}(t)$ we project onto the upper $N/2$  states. The metal-molecule coupling is turned on smoothly and slowly (in 10 fs) from preparation time.

Electronic dephasing due to vibronic couplings is incorporated by 
integrating the above equations of motion  for  an ensemble of 
initial conditions. The initial conditions are generated using 
the strategy detailed  in Sec. IIC of Ref.~\onlinecite{papaper}. 
Briefly, the starting optimal geometry of the oligomer is 
obtained by minimizing the total ground-state energy of the 
chain. Then, a normal mode analysis around this geometry is 
performed, yielding the ground state nuclear Wigner phase space 
distribution function in the harmonic approximation. By 
importance sampling this distribution,  an ensemble of lattice 
initial conditions $\{ \vect{u}^i(0), \vect{p}^i(0)\}$ is 
generated. The associated initial values for the orbitals 
$\{\ket{\epsilon^i(0)}\}$ are obtained by diagonalizing the 
electronic part of the Hamiltonian for each initial nuclear 
geometry  $\{\vect{u}^i(0)\}$. Each member $i$ of the ensemble 
defines a quantum-classical trajectory and the set is used to 
obtain ensemble averages.

Throughout this work we use the standard SSH parameters for 
PA~\cite{SSH}: $\alpha = 4.1$ eV/\AA, $K=21$ eV/\AA$^2$, 
$t_0=2.5$ eV,  $M=1349.14$ eV fs$^2$/\AA$^2$ and $a=1.22$ \AA.  
In turn, we take the molecule and leads to be weakly coupled with 
$t_\ti{coup}^2/t_\ti{lead}=0.1$ eV.

\subsection{Dynamical observables}
\label{wiresec:obs}
The vibronic dynamics is characterized by the geometric and 
spectroscopic properties defined in Sec. IID of 
Ref.~\onlinecite{papaper}.  In addition, an expression for the 
current  that depends solely  on molecular properties is required 
and is obtained in this section.

The current entering into lead $\beta=$L, R is defined by 
\be 
j_\beta = - |e| \frac{\partial p_\beta}{\partial t} 
\ee where 
$p_\beta$ is the number of electrons in lead $\beta$.
 In terms of the reduced density matrix $j_\beta$ can be expressed as
\be 
j_\beta  = - |e| \sum_{n\in \beta} \frac{\partial 
\rho_{n,n}}{\partial t}
 =  \frac{i}{\hbar} |e|  \sum_{n\in \beta, s}  \bra{\Psi(t)} [ c_{n,s}^\dagger c_{n,s}, H_\ti{elec}]\ket{\Psi(t)},
\ee where we have used Eq.~\eqref{wireeq:rdmdyn}. For the  
Hamiltonian  in \eq{eq:totelec} only the commutator with the 
lead-molecule coupling contributes to the current, \be
\label{wireeq:jtmp}
j_\beta = \frac{i}{\hbar}  |e|  \sum_{n\in \beta, s}  \bra{\Psi(t)} [ c_{n,s}^\dagger c_{n,s}, H_{\ti{S-}\beta}]\ket{\Psi(t)}.
\ee
Introducing the  projective coupling 
[\eq{wireeq:projectedcoupling}] into the above  expression, we 
have that \be
\label{wireeq:currenttmp}
\begin{split}
j_ \ti{L} & = \frac{2 |e|t_\ti{coup}}{\hbar}
\sum_{m\in \ti{S}} \text{Im}\{\mc{P}_{1,m}^\ti{S}(t)\rho_{0,m}(t) \}, \\
j_\ti{R}  & =  \frac{2 |e|t_\ti{coup}}{\hbar}
\sum_{m\in \ti{S}} \text{Im}\{\mc{P}_{N,m}^\ti{S}(t)\rho_{N+1,m}(t) \}.
\end{split}
\ee Equation~\eqref{wireeq:currenttmp} indicates that the current 
entering into the leads depends on the spatial coherences between 
the lead boundary and every site in the  molecular system.  We 
desire, however, an expression  that solely depends on molecular 
properties. For this we  exploit  \eq{wireeq:leadorbital} in its 
Markovian limit according to which, \be
\begin{split}
\langle 0 \ket{\epsilon(t)} & =
i\frac{t_\ti{coup}}{t_\ti{lead}} \sum_{m}   \mc{P}_{1,m}^{\ti{S}}(t) \langle m \ket{\epsilon_\ti{S}(t)}, \\
\langle N+1 \ket{\epsilon(t)} & =
i \frac{t_\ti{coup}}{t_\ti{lead}} \sum_{m}
\mc{P}_{N,m}^\ti{S}(t) \langle m \ket{\epsilon_\ti{S}(t)},
\end{split}
\ee so that \be
\label{wireeq:rhoboundary}
\begin{split}
\rho_{0,m}(t) &  =
-i\frac{t_\ti{coup}}{t_\ti{lead}} \sum_{n}
\mc{P}_{n,1}^{\ti{S}}(t) \rho_{n,m}(t), \\
\rho_{N+1,m}(t) &  =
-i\frac{t_\ti{coup}}{t_\ti{lead}} \sum_{n}
\mc{P}_{n,N}^{\ti{S}}(t) \rho_{n,m}(t).
\end{split}
\ee Substituting \eq{wireeq:rhoboundary} into 
\eq{wireeq:currenttmp} gives an expression for the current that 
is completely defined by molecular properties, \be
\label{wireeq:current}
j_\beta(t)  = -\frac{2|e|}{\hbar}\frac{t_\ti{coup}^2}{ 
t_\ti{lead} } \sum_{m,n} 
\text{Re}\{\mc{P}_{n_\beta,m}^{\ti{S}}(t) 
\mc{P}_{n,n_\beta}^{\ti{S}}(t) \rho_{n,m}(t)  \}, \ee as 
desired.  Another useful quantity is the amount of charge that 
has been deposited in lead $\beta$ at a given time. This is 
defined  by \be q_\beta(t) = \int_0^t j_\beta (\tau) \ud\tau. \ee
Any rectification generated by  an $\w+2\w$ pulse results in 
$q_\ti{L}-q_\ti{R}\ne 0$.

\section{Results and discussion}
\label{wiresec:results}
In Ref.~\onlinecite{papaper}  important ways in which the 
electron-vibrational couplings influence the photoinduced 
dynamics of isolated PA chains were identified. The coupling 
introduces a significant exchange of energy between vibrational 
and electronic degrees of freedom,  broadening of the electronic 
transitions,  pronounced changes in the mean single particle 
spectrum during and after the laser pulse, internal relaxation 
mechanisms and ultrafast decoherence (in less than 10 fs). By  
contrast, in rigid systems no energy is transfered between the 
electronic and nuclear degrees of freedom, and  changes in the 
electronic spectrum are only due to Stark shifts induced by the 
laser field. 

Here we analyze the extent in which the vibronic couplings affect 
the ability of $\w+2\w$ lasers to induce currents along molecular 
wires. Vibrational effects are made evident by contrasting the 
ensemble-averaged electron-vibrational evolution of the bridge 
with the dynamics of a single rigid trajectory that stays frozen  
at the  optimal geometry.  The wire is made rigid by  arbitrarily 
increasing the mass of the (CH) groups by 10$^6$. In this way the 
electron-ion interaction remains constant but the lattice moves a 
thousand times slower. Two different regimes of operation are 
considered: In \stn{wiresec:photonabsorption},  the scenario is 
applied in its usual form where the  field frequencies are tuned 
at or near an electronic resonance and electrons are promoted to 
the excited states through near-resonance photon absorption. In 
\stn{wiresec:stark} a novel mechanism is introduced   in which 
the field is kept far off resonance and excitation is done 
through the dynamic Stark effect. As shown below,  in the 
presence of vibrations the first scenario becomes  inefficient 
while the Stark shift scenario  remains  robust.

\subsection{Currents through near-resonance excitation}
\label{wiresec:photonabsorption}

\begin{table}
\caption{\label{tbl:laserparameters} Parameters and labels defining the
femtosecond laser pulses used
$E(t) = \exp(- (t-T_c)^2/T_\ti{W}^2)(\epsilon_\w \cos(\w t +
\phi_\w) + \epsilon_{2\w} \cos(2\w t + \phi_{2\w}))$. Here $I_{2\w}$ is the intensity of the $2\w$ component at maximum field strength.}
\centering
\begin{tabular}{|c| c| c | c | c| c|}
\hline
\hline
Label & $T_c$ (fs) & $T_\ti{W}$ (fs) &
$\epsilon_{2\w}$ (V \AA$^{-1}$) & $\epsilon_{\w}/\epsilon_{2\w}$  &
 $I_{2\w}$ (W cm$^{-2}$) \\
\hline
 \textbf{f1} & 900 & 300 & $8.70\times10^{-3}$ & 2.82 & $1.0\times 10^9$ \\
\textbf{f2} & 900 & 300 & $4.00\times10^{-2}$ & 2.82 & $2.1\times 10^{10}$ \\
\textbf{f3} & 50 & 10 & $8.70\times10^{-3}$ & 2.82 & $1.0\times 10^9$ \\
 \textbf{f4} & 50 & 10 & $4.00\times10^{-2}$ & 2.82 & $2.1\times 10^{10}$ \\
\hline
\hline
\end{tabular}
\end{table}

We investigate the  dynamics and currents onset by the laser 
pulses specified in Table~\ref{tbl:laserparameters} for different 
laser frequencies and relative phases noted in the text. The  
parameters chosen for the pulses are meant to encompass four 
illustrative cases: dynamics induced by weak and moderately 
strong pulses with time envelopes that are either comparable (10 
fs) or long (300 fs) compared to the typical electronic dephasing 
time.  The  laser frequencies are chosen  to be at or near  
resonance so that, in this case, near-resonance photon absorption 
is the main source of  electronic excitation.

The system consists of neutral PA wires composed of  20 sites and 
20 $\pi$ electrons ($L\approx 23$ \AA) initially in the ground 
electron-vibrational state.  An ensemble of  40,000  initial 
configurations is propagated.  This  oligomer size was selected 
because it is representative of molecules that have been  
employed in constructing molecular 
nanojunctions~\cite{salomon_review}.   The initial geometry of 
the chain consists of a perfect alternation of double and single 
bonds. The electronic structure is composed of 20 doubly occupied 
valence $\pi$ orbitals and 20 empty $\pi^\star$ states, separated 
by an energy gap of $2\Delta=1.8$ eV. A detailed account of the 
initial configuration,  is provided in Ref.~\onlinecite{papaper}.

\begin{figure}[htbp]
\centering
\psfrag{time (fs)}[][cb]{$t$ (fs)}
\psfrag{q/|e|}[][cl]{$q_\beta/|e|$}
\includegraphics[width=0.4\textwidth]{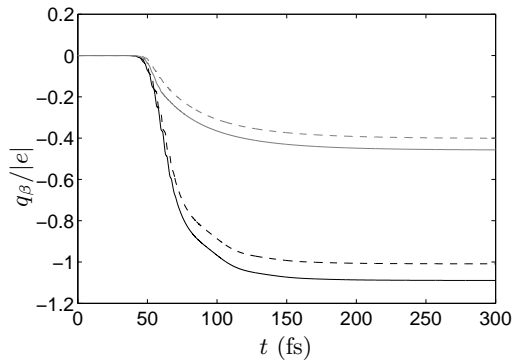}
\caption{Charge deposited in  the left (solid lines) and right (broken lines) contact  during and after photoexcitation  with pulse  \textbf{f4}.  Here $\hbar\w=1.3$ eV, $\phi_{2\w} -2\phi_\w =0$ (black lines, flexible wire; gray lines, rigid case). }
     \label{wirefig:charge10fsstrong}
\end{figure}

\begin{figure}[htbp]
\centering
\psfrag{time (fs)}[][cb]{$t$ (fs)}
\psfrag{q/|e|}[][cl]{$q_\beta/|e|$}
\includegraphics[width=0.4\textwidth]{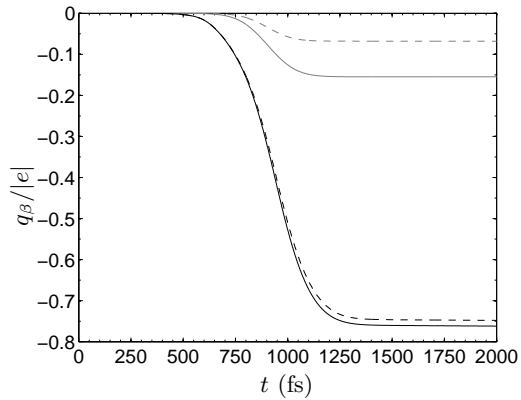}
\caption{Total charge entering the left (solid lines) and right (broken lines) lead during and after photoexcitation with pulse  \textbf{f1} with $\hbar\w=1.2$ eV, $\phi_{2\w} -2\phi_\w =0$ (gray lines, rigid wire; black line, flexible case). }
     \label{wirefig:charge300fsweak}
\end{figure}

\begin{figure}[htbp]
\centering
\psfrag{hw}[][cb]{$\hbar\w$}
\psfrag{ql-qr/e}[][]{$(q_\ti{L} - q_\ti{R})/|e|$}
\psfrag{f1}[][]{\textbf{f1}}
\psfrag{f2}[][]{\textbf{f2}}
\psfrag{f3}[][]{\textbf{f3}}
\psfrag{f4}[][]{\textbf{f4}}
\psfrag{A}[][]{A}
\psfrag{B}[][]{B}
\psfrag{C}[][]{C}
\psfrag{D}[][]{D}
\includegraphics[height=0.5\textwidth, angle=-90]{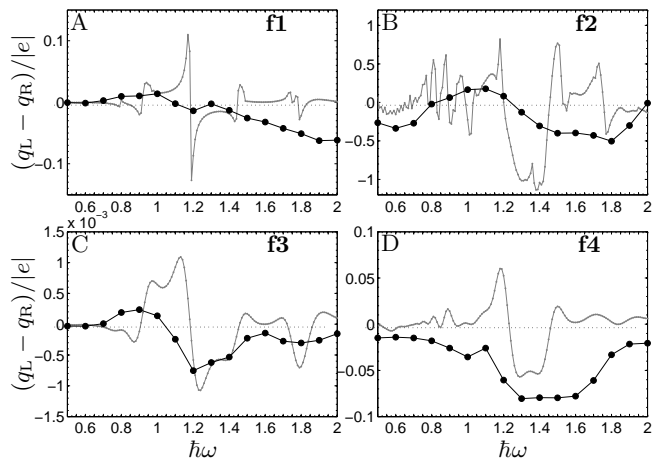}
\caption{Frequency dependence of the net difference in charge deposited by  pulse: (A) \textbf{f1}, (B) \textbf{f2},  (C) \textbf{f3}, (D) \textbf{f4} with $\phi_{2\w} - 2\phi_\w=0$. The black dots show the behavior of the flexible chain while the gray lines correspond to the rigid case. }
     \label{wirefig:diffcharge}
\end{figure}

\begin{figure}[htbp]
\centering
\psfrag{hw}[][cb]{$\hbar\w$}
\psfrag{(ql-qr)/(ql+qr)}[][]{$\eta$}
\psfrag{f1}[][]{\textbf{f1}}
\psfrag{f2}[][]{\textbf{f2}}
\psfrag{f3}[][]{\textbf{f3}}
\psfrag{f4}[][]{\textbf{f4}}
\psfrag{A}[][]{A}
\psfrag{B}[][]{B}
\psfrag{C}[][]{C}
\psfrag{D}[][]{D}
\includegraphics[height=0.5\textwidth, angle=-90]{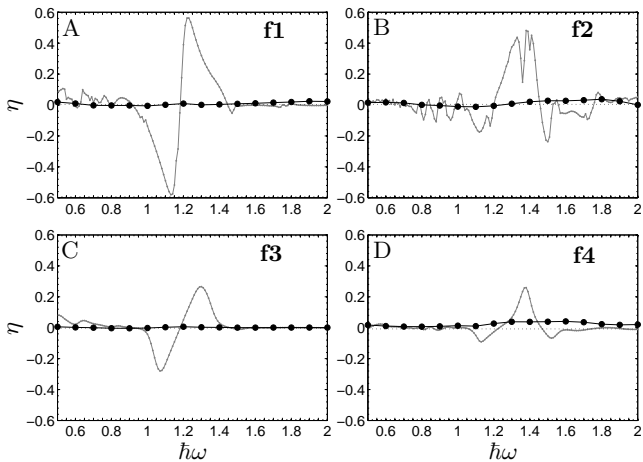}
\caption{Frequency dependence of the efficiency $\eta$ of the rectification generated by pulse: (A) \textbf{f1}, (B) \textbf{f2},  (C) \textbf{f3}, (D) \textbf{f4}, with $\phi_{2\w} - 2\phi_\w=0$. The black dots show the behavior of the flexible chain while the gray lines correspond to the rigid case. }
     \label{wirefig:quotient}
\end{figure}

\begin{figure}[htbp]
\centering
\psfrag{A}[][]{A}
\psfrag{B}[][]{B}
\psfrag{ql-qr}[][]{$(q_\ti{L} - q_\ti{R})/|e|$}
\psfrag{phi/pi}[][cb]{$(\phi_{2\w}-2\phi_\w)/\pi$}
\psfrag{ql-qr/ql+qr}[][]{$\eta$}
\includegraphics[width=0.5\textwidth]{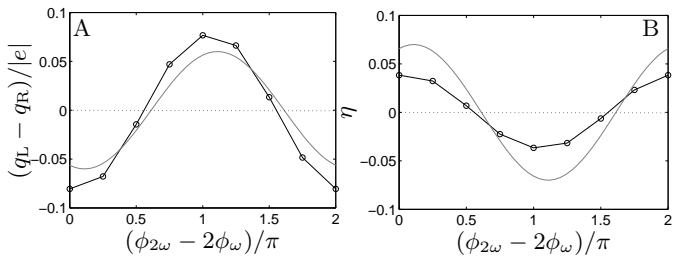}
\caption{Net transport as a function of the relative phase $\phi_{2\w}-2\phi_\w$ induced by pulse \textbf{f4} with $\hbar\w=1.3$ eV in the rigid (gray lines) and the flexible (open circles) chain. Panel (A) shows the net difference in charge deposited; (B) depicts the efficiency of the process. }
     \label{wirefig:10fs_phasedependence}
\end{figure}

We begin by presenting two concrete examples of the currents 
induced on flexible and rigid PA wires by $\w+2\w$ pulses. 
Figures~\ref{wirefig:charge10fsstrong} 
and~\ref{wirefig:charge300fsweak} show the charge deposited in 
the left and right lead during and after photoexcitation with a 
moderately strong 10 fs pulse (field \textbf{f4} in 
Table~\ref{tbl:laserparameters}) and a weak 300 fs pulse (field 
\textbf{f1}) resonantly coupling valence and conduction bands, 
respectively. Here, $\phi_{2\omega} - \phi_{\omega} = 0$.   As 
can be seen, symmetry breaking is achieved while the system is 
being driven by the laser field. After the pulse, any remaining 
electrons with energy higher than the Fermi energy are absorbed 
 symmetrically by the left and right contacts. At the particular 
frequency chosen we observe that the degree of symmetry breaking 
induced by the 10 fs pulse in the rigid and flexible chain is of 
the same order of magnitude. Note, however, that in the flexible 
wire considerably more electrons are excited across the energy 
gap than in the rigid example, due to the level broadening 
introduced by the wire's vibrations. Most of these electrons  do 
not contribute to the rectification. As discussed below, this is 
a recurrent motif when applying the near-resonance $\w+2\w$ 
scenario to flexible systems. In the case of photoexcitation with 
a 300 fs pulse, and for the parameters chosen in 
\fig{wirefig:charge300fsweak},  in rigid wires most of the   
photoexcited electrons participate in forming the dc current. In 
flexible wires, however,  most excited electrons  do not 
contribute to the net current. Although individual trajectories 
may exhibit large rectification effects, there is almost a 
perfect cancellation when the individual contributions are added 
together.  The decoherence is simply too fast to allow 
maintenance  of the rectification effect.

We now  focus on the net asymptotic current induced by all of the 
laser pulses of  Table~\ref{tbl:laserparameters}, for different 
laser frequencies. The quantities of interest are the net 
difference in charge deposited in the left and right leads  
$q_\ti{L} - q_\ti{R}$, and the overall efficiency of the process 
$\eta= (q_\ti{L}-q_\ti{R})/(q_\ti{L}+q_\ti{R})$,  where $q_\beta$ 
is the total charge deposited at lead  $\beta$.  Ideally one 
would like to find driving parameters that induce sizable 
currents in the nanojunction with high efficiency.

Figures~\ref{wirefig:diffcharge} and~\ref{wirefig:quotient} show 
$q_\ti{L}-q_\ti{R}$ and $\eta$ after excitation with pulses 
\textbf{f1}-\textbf{f4} for different laser frequencies,  with 
$\phi_{2\w}-2\phi_\w=0$. Consider first the rigid chain results.
For rigid chains the degree of control 
and the net induced currents are considerable.  The effect is 
enhanced by tuning the frequency components of the laser near one 
of the system's electronic transition frequencies.  At selected 
frequencies the process can achieve efficiencies as high as $\sim 
60\%$.  Note, however,  that for exact on-resonance driving 
frequencies the efficiencies are low due to competing  
multi-photon processes that populate the excited states but do 
not contribute to the currents. By applying stronger pulses 
(\fig{wirefig:diffcharge}B) it is possible to exploit higher 
order multiphoton processes to induce rectification, resulting in 
a complicated  frequency dependence to the control map.  When 
short pulses are employed (\fig{wirefig:diffcharge}C-D) the sharp 
features observed in the long-pulse control are no longer 
resolved. For rigid wires, higher efficiencies are achieved by 
using long weak pulses  at frequencies that minimize satellite 
photoexcitations. When employing either short or very strong 
pulses, satellite absorptions, arising  from either competing 
multiphoton processes or parasite absorption at frequencies 
contained within the bandwidth of the pulse, are unavoidable.

The  inclusion  of  vibrations  in  the dynamics substantially
modifies     the     physical    picture.    As    shown    in
Figs.~\ref{wirefig:diffcharge}     and~\ref{wirefig:quotient},
lattice  fluctuations  smear  out  the  resonances observed in
rigid  wires  resulting  in  control  maps with broad features
only.  Although  the  scenario  is  successful  in  generating
phase       controllable       currents       (see,      e.g.,
\fig{wirefig:10fs_phasedependence})  with  magnitudes that are
often  larger  than  the  ones  observed  in  rigid wires, the
efficiency   of   the   process   is   invariably   very   low
irrespective  of  the  driving  frequency,  laser intensity or
bandwidth     (\fig{wirefig:quotient}).    That    is,    most
photoexcited  electrons  do  not  participate  of the currents
and  the  enhancement  of the effect is due to increased light
absorption  of  the  flexible wire relative to the rigid case.
The  highest  observed  efficiencies,  obtained for moderately
strong                       short                      pulses
(Figs.~\ref{wirefig:diffcharge}-\ref{wirefig:quotient}D),
are  merely  $\eta\approx  0.04$.  These  results  clarify the
previously   unexplained   origin   for   the  enhancement  of
currents  by  increasing  vibronic  coupling strength observed
in Ref.~\cite{lehmann_1vs2}.

\subsection{Currents through the dynamic Stark effect}
\label{wiresec:stark}

Thus far our approach has been to excite electrons across the energy gap of the wire by tuning the frequencies of the field near a resonance of the system. We have seen, however, that this 
technique is fragile to decoherence processes since it relies on creating coherent superposition states.  Lattice fluctuations  make the rectification  extremely inefficient and, in the best of cases, only about $4\%$ of the  electrons that are photoexcited  participate in the net current. Hence, in order to induce sizable currents considerable energy from the field 
must be dumped into the nanojunction, compromising its structural integrity. Further, employing faster or stronger pulses is not very helpful since they introduce other 
undesired satellite channels in the control and do not appreciably overcome the deleterious effects introduced by the vibrations.

In this section  we introduce an alternative mechanism that is remarkably robust to electron-vibrational couplings, survives in the presence of decoherence and thermal effects, and is able to induce large currents in molecular wires with  efficiencies $> 90\%$. Instead of promoting electrons to the conduction band through near-resonance photon absorption, we work far off resonance and employ Stark shifts to nonadiabatically couple the  ground and excited electronic states. Phase-controllable symmetry breaking is achieved by exploiting the difference in the intensity of $\w+2\w$ fields for positive and negative amplitudes.

Although this mechanism is applicable in medium sized chains ($N\approx 20$)  it is best suited for  long oligomers. Longer wires have a smaller energy gap, facilitating  the coupling of ground and excited states  through Stark shifts before photon absorption becomes dominant.  For this reason we consider perfectly dimerized neutral PA oligomers with $N=100$ sites ($L\approx 120.8$ \AA) originally in the ground state configuration, and propagate an ensemble of 1000 initial configurations.  The electronic structure   consists of $N/2$ doubly occupied $\pi$ (valence) orbitals
and $N/2$ empty $\pi^\star$ (conduction band) states,  separated by an
energy gap of $2\Delta=1.3$ eV which is significantly   smaller than the 1.8 eV observed by   20 site oligomers.
The average total energy of the ensemble is  $\sim 4.4$
eV higher than its rigid counterpart due to zero-point fluctuations of the
lattice in the ground electronic surface.

\subsubsection{Complete control}

We now illustrate the essentials of the phenomenon and show how under realistic conditions  one can achieve almost complete control over the currents even in the presence of significant electron-phonon couplings. Consider the electron-vibrational dynamics of the wire under the influence of an $\w+2\w$ field of the form
\be
\begin{split}
\label{wireeq:nafield2}
E(t) & = \sum_{n=1}^{2}  \epsilon_{n\w}\cos(n\w t+\phi_{n\w})
\times \\
& \left\{ \begin{array}{ll}
\exp{(-{(t -t_\ti{on})^2}/{T_w^2})} & \textrm{for $t \le t_\ti{on}$} \\
1 & \textrm{for $ t_\ti{on} < t < t_\ti{off}$} \\
\exp{(-{(t -t_\ti{off})^2}/{T_w^2})} & \textrm{for $t \ge t_\ti{off}$}
\end{array} \right..
\end{split}
\ee
 The field is smoothly turned on and off in  $T_w=100$ fs ($t_\ti{on} =300$ fs; $t_\ti{off} = 700$ fs) and has  constant amplitude for 400 fs.   The frequency $\hbar\w=0.2\Delta= 0.13$ eV is chosen far off-resonance from the system's interband transition frequencies so that Stark shifts, and not photon absorption, dominate the dynamics. The field amplitude used is $\epsilon_{2\w} = 6.1\times 10^{-3}$ V \AA$^{-1}$ with $\epsilon_{\w} = 2\epsilon_{2\w}$, which corresponds to an intensity $I_{2\w} \sim 5\times 10^{8}$ W cm$^{-2}$.

\begin{figure}[htbp]
\centering
\psfrag{t (fs)}[][]{$t$ (fs)}
\psfrag{E(t)}[][]{$E(t)$ (V \AA$^{-1}$)}
\psfrag{jbeta}[][]{$j_\beta/|e|$ (fs$^{-1}$)}
\psfrag{jl}[][]{$j_\ti{L}$}
\psfrag{jr}[][]{$j_\ti{R}$}
\psfrag{energy}[][]{$\mean{\epsilon_\gamma}$ (eV)}
\psfrag{A}[][]{A}
\psfrag{B}[][]{B}
\psfrag{C}[][]{C}
\includegraphics[width=0.5\textwidth, angle=0]{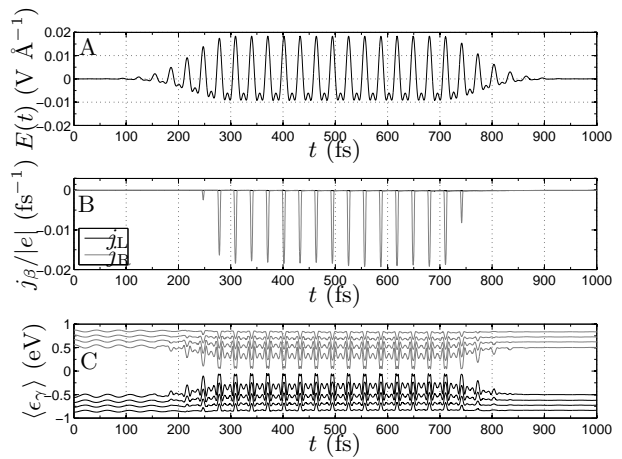}
\caption{Time dependence of (A) the electric field, (B) the current entering the left and right leads, and (C) the instantaneous light-dressed orbital energies for states near the energy gap for a 100-site flexible wire under the influence of the field in \eq{wireeq:nafield2}  with $\phi_{2\w}-2\phi_\w=0$. Note the bursts of charge deposited in the right lead when the field bridges the energy gap. Here $j_\ti{L}$ is so small that is barely visible.}
     \label{wirefig:flexnotgaussian_0}
\end{figure}

\begin{figure}[htbp]
\centering
\psfrag{t (fs)}[][]{$t$ (fs)}
\psfrag{E(t)}[][]{$E(t)$ (V \AA$^{-1}$)}
\psfrag{jbeta}[][]{$j_\beta/|e|$ (fs$^{-1}$)}
\psfrag{jl }[][]{$j_\ti{L}$}
\psfrag{jr }[][]{$j_\ti{R}$}
\psfrag{energy}[][]{$\mean{\epsilon_\gamma}$ (eV)}
\psfrag{A}[][]{A}
\psfrag{B}[][]{B}
\psfrag{C}[][]{C}
\includegraphics[width=0.5\textwidth, angle=0]{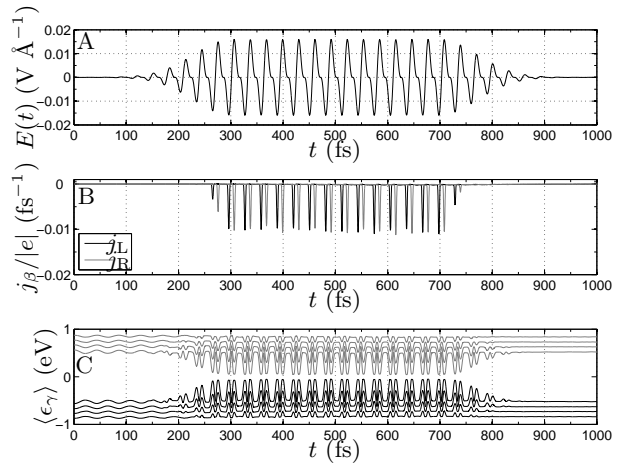}
\caption{Time dependence of (A) the electric field; (B) the current entering the left and right leads and; (C) the instantaneous light-dressed orbital energies for states near the energy gap for a 100-site flexible wire under the influence of the field in \eq{wireeq:nafield2},  with $\phi_{2\w}-2\phi_\w=\pi/2$.  Note that in this case the bursts of charge deposited alternate between the left and right contact.}
     \label{wirefig:flexnotgaussian_90}
\end{figure}

\begin{figure}[htbp]
\centering
\psfrag{phi/pi}[][]{$(\phi_{2\w}-2\phi_\w)/\pi$}
\psfrag{ql-qr}[][]{$(q_\ti{L}-q_\ti{R})/|e|$}
\psfrag{eta}[][]{$\eta$}
\psfrag{A}[][]{A}
\psfrag{B}[][]{B}
\includegraphics[width=0.5\textwidth, angle=0]{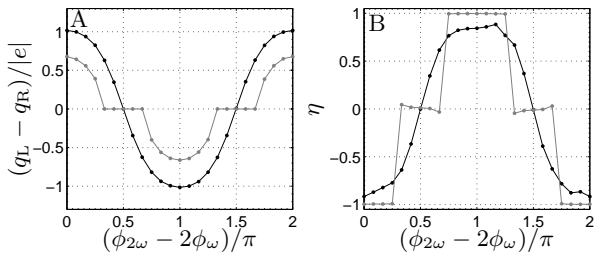}
\caption{Phase dependence of (A) the net rectification, and (B) efficiency of the process when the field in \eq{wireeq:nafield2} is applied to flexible and rigid 100-site PA wires (black dots, flexible wire; gray dots, rigid case). Note the assistance of phonons in the rectification.}
     \label{wirefig:notgaussian_phase_dependence}
\end{figure}

 Figure~\ref{wirefig:flexnotgaussian_0} shows the field, the currents, and the light-dressed electronic structure of the flexible chain, averaged over all trajectories, when the relative phase of the  pulse is $\phi_{2\w}-2\phi_\w=0$. Due to the highly polarizable nature of $\pi$-conjugated systems, the single-particle spectrum  displays  considerable  Stark shifts,  even at moderate field intensities; the effect being  stronger for states near the edges of the valence and conduction band. In this way, the dynamic Stark effect  effectively reduces the energy gap of the oligomer, causing frequent crossings between the  ground and excited electronic states in individual trajectories. Since the wire's ground state is nondegenerate and of definite parity, the lowest-order contribution to the Stark effect is quadratic in the field. Hence, when $|E(t)|$ is maximum the energy gap acquires  acquires its minimum value. At the crossing times population is transferred from the valence to the conduction band and bursts of charge are deposited in the leads.

Note that for $\phi_{2\w}-2\phi_\w=0$ almost all excited electrons are deposited in the right contact only. Symmetry breaking arises due to the difference in the maximum $|E(t)|$ for positive and negative amplitudes exhibited by the field. Even when $E(t)$ has a zero temporal mean, it consists of narrow peaks with large $|E(t)|$ for positive amplitudes, and shallow and broad features when $E(t)$ is negative (\fig{wirefig:flexnotgaussian_0}A). For this reason, the Stark effect is only sufficiently strong to close the energy gap when the field has a positive amplitude. Thus, transfer of population to the conduction band and absorption of electrons by the leads always occur when the laser is pointing at a particular and the same direction, in this way inducing directed transport in the system.

The phenomenon depends intimately on the relative phase. For instance, for the case of  $\phi_{2\w}-2\phi_\w=\pi/2$  shown in \fig{wirefig:flexnotgaussian_90}  the electric field exhibits equal intensity for positive and negative amplitudes. Since the field changes sign between consecutive interband couplings, the bursts of charge  deposited   alternate between the left and right contact, and no net current is induced.

Figure~\ref{wirefig:notgaussian_phase_dependence} shows the net difference in charge deposited in the left and right leads after the pulse is over for different laser phases, as well as the efficiency of the process. For comparison purposes the plot also includes results obtained in an equivalent but rigid system. We first note that the mechanism is robust to decoherence effects due to coupling to the vibrational degrees of freedom as well as satellite contributions due to parasite multiphoton absorption. In fact, 90\% of the excited electrons can participate in the net current. This should be contrasted to the extremely low efficiencies achieved before  through the multi-photon absorption mechanism. Further, the sign and magnitude of the effect can be manipulated by varying the laser phases. By changing the relative phase by $\pi$ the magnitude of the effect stays the same but the direction of the rectification is reversed.

In the flexible wire the rectification exhibits an almost sinusoidal dependence on $\phi_{2\w}-2\phi_\w$. By contrast, in rigid wires for  certain range of  phases no currents are induced since the maximum field amplitude is not large enough to  couple valence and conduction bands. Hence, in this range the currents observed in the flexible wire are phonon-assisted; i.e. the level broadening introduced by the vibrations permit the nonadiabatic coupling. In fact, the currents observed in the flexible case are always larger than in the rigid example. However, in rigid wires the mechanism can exhibit perfect efficiencies.  The mechanism is also expected to be resilient to thermal effects. Increasing the temperature will introduce additional  broadening of the electronic levels and, as pointed above, the Stark rectification  is robust to this class of effects.

 \subsubsection{Possible satellite contributions}

\begin{figure*}[htbp]
\centering
\psfrag{E(t)}[][]{$E(t)$ (V \AA$^{-1}$)}
\psfrag{t (fs)}[][]{$t$ (fs)}
\psfrag{jbeta}[][]{$j_\beta/|e|$ (fs$^{-1}$)}
\psfrag{JR}[][]{$j_\ti{R}$}
\psfrag{JL}[][]{$j_\ti{L}$}
\psfrag{A}[][]{A}
\psfrag{B}[][]{B}
\includegraphics[width=0.8\textwidth, angle=0]{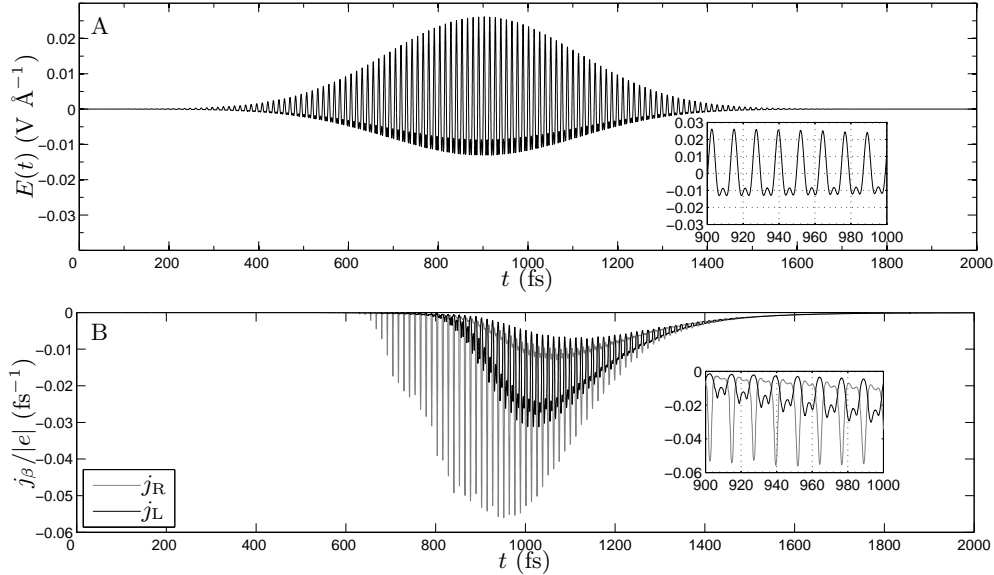}
\caption{Time dependence of (A) the electric field, and (B) the current entering the left and right leads  for a 100-site flexible wire under the influence of the field in \eq{wireeq:nafield}  with $\phi_{2\w}-2\phi_\w=0$. The insets detail the field and currents from $t=900$ fs to $1000$ fs. }
     \label{wirefig:100flexcomposed}
\end{figure*}

\begin{figure}[htbp]
\centering
\psfrag{t (fs)}[][]{$t$ (fs)}
\psfrag{qbeta}[][]{$q_\beta/|e|$}
\psfrag{qr}[][]{$q_\ti{R}$}
\psfrag{ql}[][]{$q_\ti{L}$}
\includegraphics[width=0.5\textwidth, angle=0]{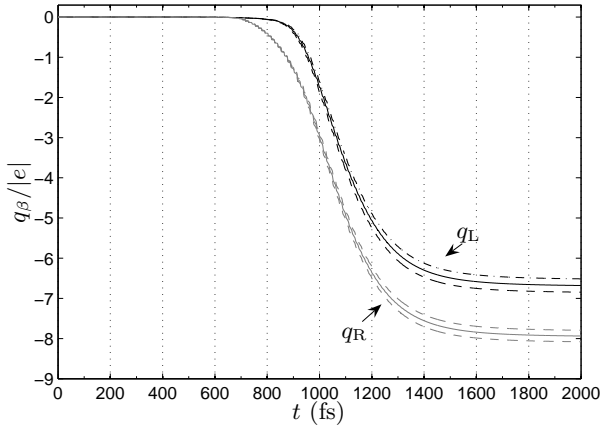}
\caption{Total charge deposited in the left $q_\ti{L}$ (black line) and right $q_\ti{R}$ (gray line) contact during the dynamics of a 100-site flexible wire under the influence of the field in \eq{wireeq:nafield},  with $\phi_{2\w}-2\phi_\w=0$. The dotted lines provide an estimate of the sampling errors. }
     \label{wirefig:100flexcharge}
\end{figure}

The simulations presented above exemplify how almost complete control over the electronic dynamics in the presence of vibronic couplings can be exerted in molecular wires. However, it is interesting to point out possible sources of satellite contributions that reduce the efficiency of this rectification mechanism. For definitiveness consider the wire's dynamics under the influence of an $\w+2\w$ Gaussian pulse of the form
\be
\label{wireeq:nafield}
E(t) = e^{-(t-T_c)^2/T_W^2} \sum_{n=1}^{2} \epsilon_{n\w} \cos(n\w t + \phi_{n\w}),
\ee
 centered at $T_c=900$ fs, with  temporal width $T_W=300$ fs, and amplitudes $\epsilon_{2\w} = 8.7\times 10^{-3}$ V \AA$^{-1}$ with $\epsilon_{\w}=2\epsilon_{2\w}$, which corresponds to an intensity of $I_{2\w}=10^9$ W cm$^{-2}$, and driving frequency    $\hbar\w= 0.46\Delta=0.3$ eV. With respect to the field in \eq{wireeq:nafield2}, this laser profile has a higher maximum intensity and blue shifted central frequencies. Figure~\ref{wirefig:100flexcomposed} shows the currents onset by this laser pulse when the relative phase is $\phi_{2\w}-2\phi_\w=0$, and \fig{wirefig:100flexcharge} the total charge deposited during the dynamics. The field generates net currents along the system, with the right contact accepting 1.25 more electrons than the left one. However, in this case only 9$\%$ of the excited electrons participate in the net current.

 There are two sources of satellite contributions that reduce the efficiency of the scenario. 
The first is that some interband coupling occurs when the amplitude of the field is negative, contrary to the case described in the previous section. That is, the shallow negative amplitude of the field is sufficient to close the energy gap in some trajectories and deposit electrons in the left lead. 
Second, since the field frequency is larger than the one in \eq{wireeq:nafield2} there is considerable excitation originating from photon absorption and not from Stark shifts. This leads to a background signal in $j_\beta(t)$ that does not contribute appreciably to the rectification and reduces the efficiency of the process. It is not difficult to suppress these two satellite effects by changing field parameters. Parasite multiphoton absorption can be avoided by further detuning the frequency of the field from the interband transition frequencies. Undesired closing of the energy gap can be circumvented by employing a laser profile with controlled and reduced intensity.
Highly efficient current generation, such as that seen in the previous example, results.

\section{Conclusions}
\label{wiresec:conclusions}

Here we have studied the possibility of inducing ultrafast currents using $\w+2\w$ fields  along molecular wires with significant electron-vibrational couplings.  Two possible mechanism for laser-inducing rectification were identified. In the first one, electrons are excited across the energy gap of the wire by tuning the frequencies of the field near an electronic resonance of the system. By applying lasers with both $\w$ and $2\w$ components asymmetry is generated in the momentum distribution of the photoexcited electrons. In the second mechanism, the $\w+2\w$ field is kept far off-resonance and population is transferred to the excited states by means of Stark shifts that bridge the energy gap of the oligomer and, in this way, couple ground and excited electronic states. Symmetry breaking is achieved by exploiting the difference in the field's intensity  for positive and negative  amplitudes. This permits transferring population across the energy gap only when the field is pointing at a particular direction, thereby breaking the spatial symmetry of the system. In both cases the magnitude and direction of the net currents is controllable by varying the relative phase of the two frequency components.

The vibronic couplings in the wire have very different effects in these two cases. In the first one,   ultrafast decoherence processes induced by the wire's vibrations make the rectification effect extremely inefficient.  While in rigid wires the maximum efficiency observed is $\sim 60\%$, in flexible chains less than  $4\%$ of the photoexcited electrons participate in the current. This trend holds  irrespective of the driving frequency, field strength or width.  At selected frequencies the vibrations can enhance the magnitude of the effect, but not the efficiency, due to increased light absorption through level broadening. This observation sheds light on the previously unexplained observations reported in Ref.~\cite{lehmann_1vs2} where the magnitude of the photoinduced currents was shown to increase with the strenght of the vibronic couplings. Note, however, that in order to generate sizable currents in this regime  a considerable 
number of electrons must be excited, undermining the structural stability of the system.

In contrast to the above mechanism, the off-resonance  Stark mechanism introduced here is impervious to vibronic couplings. Even when the lattice dynamics may introduce parasite multiphoton absorption or undesired closing of the energy gap, these problems are easily  overcome  by red shifting the driving frequency and using pulses  of lowered  and controlled intensity. Simulations show that the mechanism can actually induce currents with efficiencies as high as $90\%$, in spite of the significant electron-phonon couplings.

We note that the actual implementation of these schemes poses substantial experimental challenges. The laser beam needs to be  focused on the wire and  the setup must control possible side effects that may arise from the illumination, such as  heating, or processes involving excitation of the metal surface. Recent experimental developments~\cite{wusw, lipsonmichal, aeschlimann} suggest that this may be possible. We hope that emerging applications, like the one presented here,  stimulate further experimental progress.

Last, from the simulations presented herein and a recent experiment~\cite{sussman}, it is clear that  the dynamic Stark effect is a promising tool to exert laser control that deserves further exploration.  The success of this approach relies on the fact that it does not exploit  the fragile properties of  superposition states.

\bibliography{609825JCP}

\begin{thebibliography}{44}
\expandafter\ifx\csname natexlab\endcsname\relax\def\natexlab#1{#1}\fi
\expandafter\ifx\csname bibnamefont\endcsname\relax
  \def\bibnamefont#1{#1}\fi
\expandafter\ifx\csname bibfnamefont\endcsname\relax
  \def\bibfnamefont#1{#1}\fi
\expandafter\ifx\csname citenamefont\endcsname\relax
  \def\citenamefont#1{#1}\fi
\expandafter\ifx\csname url\endcsname\relax
  \def\url#1{\texttt{#1}}\fi
\expandafter\ifx\csname urlprefix\endcsname\relax\def\urlprefix{URL }\fi
\providecommand{\bibinfo}[2]{#2}
\providecommand{\eprint}[2][]{\url{#2}}

\bibitem[{\citenamefont{Lindsay and Ratner}(2007)}]{lindsay_review}
\bibinfo{author}{\bibfnamefont{S.~M.} \bibnamefont{Lindsay}} \bibnamefont{and}
  \bibinfo{author}{\bibfnamefont{M.~A.} \bibnamefont{Ratner}},
  \bibinfo{journal}{Adv. Mat.} \textbf{\bibinfo{volume}{19}},
  \bibinfo{pages}{23} (\bibinfo{year}{2007}).

\bibitem[{\citenamefont{Nitzan and Ratner}(2003)}]{nitzan_review}
\bibinfo{author}{\bibfnamefont{A.}~\bibnamefont{Nitzan}} \bibnamefont{and}
  \bibinfo{author}{\bibfnamefont{M.~A.} \bibnamefont{Ratner}},
  \bibinfo{journal}{Science} \textbf{\bibinfo{volume}{300}},
  \bibinfo{pages}{1384} (\bibinfo{year}{2003}).

\bibitem[{\citenamefont{Heath and Ratner}(2003)}]{ratner_review}
\bibinfo{author}{\bibfnamefont{J.~R.} \bibnamefont{Heath}} \bibnamefont{and}
  \bibinfo{author}{\bibfnamefont{M.~A.} \bibnamefont{Ratner}},
  \bibinfo{journal}{Phys. Today} \textbf{\bibinfo{volume}{56}},
  \bibinfo{pages}{43} (\bibinfo{year}{2003}).

\bibitem[{\citenamefont{Adams et~al.}(2003)\citenamefont{Adams, Brus, Chidsey,
  Creager, Creutz, Kagan, Kamat, Lieberman, Lindsay, Marcus
  et~al.}}]{adams_review}
\bibinfo{author}{\bibfnamefont{D.~M.} \bibnamefont{Adams}},
  \bibinfo{author}{\bibfnamefont{L.}~\bibnamefont{Brus}},
  \bibinfo{author}{\bibfnamefont{C.~E.~D.} \bibnamefont{Chidsey}},
  \bibinfo{author}{\bibfnamefont{S.}~\bibnamefont{Creager}},
  \bibinfo{author}{\bibfnamefont{C.}~\bibnamefont{Creutz}},
  \bibinfo{author}{\bibfnamefont{C.~R.} \bibnamefont{Kagan}},
  \bibinfo{author}{\bibfnamefont{P.~V.} \bibnamefont{Kamat}},
  \bibinfo{author}{\bibfnamefont{M.}~\bibnamefont{Lieberman}},
  \bibinfo{author}{\bibfnamefont{S.}~\bibnamefont{Lindsay}},
  \bibinfo{author}{\bibfnamefont{R.~A.} \bibnamefont{Marcus}},
  \bibnamefont{et~al.}, \bibinfo{journal}{J. Phys. Chem. B}
  \textbf{\bibinfo{volume}{107}}, \bibinfo{pages}{6668} (\bibinfo{year}{2003}).

\bibitem[{\citenamefont{Salomon et~al.}(2003)\citenamefont{Salomon, Cahen,
  Lindsay, Tomfohr, Engelkes, and Frisbie}}]{salomon_review}
\bibinfo{author}{\bibfnamefont{A.}~\bibnamefont{Salomon}},
  \bibinfo{author}{\bibfnamefont{D.}~\bibnamefont{Cahen}},
  \bibinfo{author}{\bibfnamefont{S.}~\bibnamefont{Lindsay}},
  \bibinfo{author}{\bibfnamefont{J.}~\bibnamefont{Tomfohr}},
  \bibinfo{author}{\bibfnamefont{V.~B.} \bibnamefont{Engelkes}},
  \bibnamefont{and} \bibinfo{author}{\bibfnamefont{C.~D.}
  \bibnamefont{Frisbie}}, \bibinfo{journal}{Adv. Mater.}
  \textbf{\bibinfo{volume}{15}}, \bibinfo{pages}{1881} (\bibinfo{year}{2003}).

\bibitem[{\citenamefont{Joachim and Ratner}(2005)}]{joachim_review}
\bibinfo{author}{\bibfnamefont{C.}~\bibnamefont{Joachim}} \bibnamefont{and}
  \bibinfo{author}{\bibfnamefont{M.~A.} \bibnamefont{Ratner}},
  \bibinfo{journal}{Proc. Natl. Acad. Sci. USA} \textbf{\bibinfo{volume}{102}},
  \bibinfo{pages}{8801} (\bibinfo{year}{2005}).

\bibitem[{\citenamefont{Nitzan}(2001)}]{nitzan}
\bibinfo{author}{\bibfnamefont{A.}~\bibnamefont{Nitzan}},
  \bibinfo{journal}{Annu. Rev. Phys. Chem.} \textbf{\bibinfo{volume}{52}},
  \bibinfo{pages}{681} (\bibinfo{year}{2001}).

\bibitem[{\citenamefont{Shapiro and Brumer}(2003)}]{paul}
\bibinfo{author}{\bibfnamefont{M.}~\bibnamefont{Shapiro}} \bibnamefont{and}
  \bibinfo{author}{\bibfnamefont{P.}~\bibnamefont{Brumer}},
  \emph{\bibinfo{title}{Principles of the Quantum Control of Molecular
  Processes}} (\bibinfo{publisher}{John Wiley \& Sons}, \bibinfo{address}{New
  York}, \bibinfo{year}{2003}).

\bibitem[{\citenamefont{Franco and Brumer}(2006)}]{francoprl}
\bibinfo{author}{\bibfnamefont{I.}~\bibnamefont{Franco}} \bibnamefont{and}
  \bibinfo{author}{\bibfnamefont{P.}~\bibnamefont{Brumer}},
  \bibinfo{journal}{Phys. Rev. Lett.} \textbf{\bibinfo{volume}{97}},
  \bibinfo{pages}{040402} (\bibinfo{year}{2006}).

\bibitem[{\citenamefont{Cizek et~al.}(2004)\citenamefont{Cizek, Thoss, and
  Domcke}}]{cizek}
\bibinfo{author}{\bibfnamefont{M.}~\bibnamefont{Cizek}},
  \bibinfo{author}{\bibfnamefont{M.}~\bibnamefont{Thoss}}, \bibnamefont{and}
  \bibinfo{author}{\bibfnamefont{W.}~\bibnamefont{Domcke}},
  \bibinfo{journal}{Phys. Rev. B} \textbf{\bibinfo{volume}{70}},
  \bibinfo{pages}{125406} (\bibinfo{year}{2004}).

\bibitem[{\citenamefont{Emberly and Kirczenow}(2000)}]{emberly}
\bibinfo{author}{\bibfnamefont{E.~G.} \bibnamefont{Emberly}} \bibnamefont{and}
  \bibinfo{author}{\bibfnamefont{G.}~\bibnamefont{Kirczenow}},
  \bibinfo{journal}{Phys. Rev. B} \textbf{\bibinfo{volume}{61}},
  \bibinfo{pages}{5740} (\bibinfo{year}{2000}).

\bibitem[{\citenamefont{Galperin et~al.}(2004)\citenamefont{Galperin, Ratner,
  and Nitzan}}]{galperin_vibrations1}
\bibinfo{author}{\bibfnamefont{M.}~\bibnamefont{Galperin}},
  \bibinfo{author}{\bibfnamefont{M.~A.} \bibnamefont{Ratner}},
  \bibnamefont{and} \bibinfo{author}{\bibfnamefont{A.}~\bibnamefont{Nitzan}},
  \bibinfo{journal}{J. Chem. Phys.} \textbf{\bibinfo{volume}{121}},
  \bibinfo{pages}{11965} (\bibinfo{year}{2004}).

\bibitem[{\citenamefont{Galperin et~al.}(2006)\citenamefont{Galperin, Nitzan,
  and Ratner}}]{galperin_vibrations2}
\bibinfo{author}{\bibfnamefont{M.}~\bibnamefont{Galperin}},
  \bibinfo{author}{\bibfnamefont{A.}~\bibnamefont{Nitzan}}, \bibnamefont{and}
  \bibinfo{author}{\bibfnamefont{M.~A.} \bibnamefont{Ratner}},
  \bibinfo{journal}{Phys. Rev. B} \textbf{\bibinfo{volume}{74}},
  \bibinfo{pages}{075326} (\bibinfo{year}{2006}).

\bibitem[{\citenamefont{May}(2002)}]{may}
\bibinfo{author}{\bibfnamefont{V.}~\bibnamefont{May}}, \bibinfo{journal}{Phys.
  Rev. B} \textbf{\bibinfo{volume}{66}}, \bibinfo{pages}{245411}
  (\bibinfo{year}{2002}).

\bibitem[{\citenamefont{Ness and Fisher}(1999)}]{ness1}
\bibinfo{author}{\bibfnamefont{H.}~\bibnamefont{Ness}} \bibnamefont{and}
  \bibinfo{author}{\bibfnamefont{A.~J.} \bibnamefont{Fisher}},
  \bibinfo{journal}{Phys. Rev. Lett.} \textbf{\bibinfo{volume}{83}},
  \bibinfo{pages}{452} (\bibinfo{year}{1999}).

\bibitem[{\citenamefont{Ness et~al.}(2001)\citenamefont{Ness, Shevlin, and
  Fisher}}]{ness2}
\bibinfo{author}{\bibfnamefont{H.}~\bibnamefont{Ness}},
  \bibinfo{author}{\bibfnamefont{S.~A.} \bibnamefont{Shevlin}},
  \bibnamefont{and} \bibinfo{author}{\bibfnamefont{A.~J.}
  \bibnamefont{Fisher}}, \bibinfo{journal}{Phys. Rev. B}
  \textbf{\bibinfo{volume}{63}}, \bibinfo{pages}{125422}
  (\bibinfo{year}{2001}).

\bibitem[{\citenamefont{Troisi et~al.}(2003)\citenamefont{Troisi, Ratner, and
  Nitzan}}]{troisi}
\bibinfo{author}{\bibfnamefont{A.}~\bibnamefont{Troisi}},
  \bibinfo{author}{\bibfnamefont{M.~A.} \bibnamefont{Ratner}},
  \bibnamefont{and} \bibinfo{author}{\bibfnamefont{A.}~\bibnamefont{Nitzan}},
  \bibinfo{journal}{J. Chem. Phys.} \textbf{\bibinfo{volume}{118}},
  \bibinfo{pages}{6072} (\bibinfo{year}{2003}).

\bibitem[{\citenamefont{Heeger}(2001)}]{HeegerAJ:NobLSm}
\bibinfo{author}{\bibfnamefont{A.~J.} \bibnamefont{Heeger}},
  \bibinfo{journal}{Rev. Mod. Phys.} \textbf{\bibinfo{volume}{73}},
  \bibinfo{pages}{681} (\bibinfo{year}{2001}).

\bibitem[{\citenamefont{Heeger et~al.}(1988)\citenamefont{Heeger, Kivelson,
  Schrieffer, and Su}}]{SSH}
\bibinfo{author}{\bibfnamefont{A.~J.} \bibnamefont{Heeger}},
  \bibinfo{author}{\bibfnamefont{S.}~\bibnamefont{Kivelson}},
  \bibinfo{author}{\bibfnamefont{J.~R.} \bibnamefont{Schrieffer}},
  \bibnamefont{and} \bibinfo{author}{\bibfnamefont{W.~P.} \bibnamefont{Su}},
  \bibinfo{journal}{Rev. Mod. Phys.} \textbf{\bibinfo{volume}{60}},
  \bibinfo{pages}{781} (\bibinfo{year}{1988}).

\bibitem[{\citenamefont{Franco et~al.}(2007{\natexlab{a}})\citenamefont{Franco,
  Shapiro, and Brumer}}]{papaper}
\bibinfo{author}{\bibfnamefont{I.}~\bibnamefont{Franco}},
  \bibinfo{author}{\bibfnamefont{M.}~\bibnamefont{Shapiro}}, \bibnamefont{and}
  \bibinfo{author}{\bibfnamefont{P.}~\bibnamefont{Brumer}}
  (\bibinfo{year}{2007}{\natexlab{a}}), \bibinfo{note}{previous paper}.

\bibitem[{\citenamefont{Dulic et~al.}(2003)\citenamefont{Dulic, van~der Molen,
  Kudernac, Jonkman, de~Jong, Bowden, van Esch, Feringa, and van
  Wees}}]{switch1}
\bibinfo{author}{\bibfnamefont{D.}~\bibnamefont{Dulic}},
  \bibinfo{author}{\bibfnamefont{S.~J.} \bibnamefont{van~der Molen}},
  \bibinfo{author}{\bibfnamefont{T.}~\bibnamefont{Kudernac}},
  \bibinfo{author}{\bibfnamefont{H.~T.} \bibnamefont{Jonkman}},
  \bibinfo{author}{\bibfnamefont{J.~J.~D.} \bibnamefont{de~Jong}},
  \bibinfo{author}{\bibfnamefont{T.~N.} \bibnamefont{Bowden}},
  \bibinfo{author}{\bibfnamefont{J.}~\bibnamefont{van Esch}},
  \bibinfo{author}{\bibfnamefont{B.~L.} \bibnamefont{Feringa}},
  \bibnamefont{and} \bibinfo{author}{\bibfnamefont{B.~J.} \bibnamefont{van
  Wees}}, \bibinfo{journal}{Phys. Rev. Lett.} \textbf{\bibinfo{volume}{91}},
  \bibinfo{pages}{207402} (\bibinfo{year}{2003}).

\bibitem[{\citenamefont{Yasutomi et~al.}(2004)\citenamefont{Yasutomi, Morita,
  Imanishi, and Kimura}}]{switch2}
\bibinfo{author}{\bibfnamefont{S.}~\bibnamefont{Yasutomi}},
  \bibinfo{author}{\bibfnamefont{T.}~\bibnamefont{Morita}},
  \bibinfo{author}{\bibfnamefont{Y.}~\bibnamefont{Imanishi}}, \bibnamefont{and}
  \bibinfo{author}{\bibfnamefont{S.}~\bibnamefont{Kimura}},
  \bibinfo{journal}{Science} \textbf{\bibinfo{volume}{304}},
  \bibinfo{pages}{1944} (\bibinfo{year}{2004}).

\bibitem[{\citenamefont{Zhang et~al.}(2004)\citenamefont{Zhang, Du, Cheng,
  Zhang, Roitberg, and Krause}}]{switchtheory}
\bibinfo{author}{\bibfnamefont{C.}~\bibnamefont{Zhang}},
  \bibinfo{author}{\bibfnamefont{M.~H.} \bibnamefont{Du}},
  \bibinfo{author}{\bibfnamefont{H.~P.} \bibnamefont{Cheng}},
  \bibinfo{author}{\bibfnamefont{X.~G.} \bibnamefont{Zhang}},
  \bibinfo{author}{\bibfnamefont{A.~E.} \bibnamefont{Roitberg}},
  \bibnamefont{and} \bibinfo{author}{\bibfnamefont{J.~L.}
  \bibnamefont{Krause}}, \bibinfo{journal}{Phys. Rev. Lett.}
  \textbf{\bibinfo{volume}{92}}, \bibinfo{pages}{158301}
  (\bibinfo{year}{2004}).

\bibitem[{\citenamefont{Tikhonov
  et~al.}(2002{\natexlab{a}})\citenamefont{Tikhonov, Coalson, and
  Dahnovsky}}]{tikhonov_1}
\bibinfo{author}{\bibfnamefont{A.}~\bibnamefont{Tikhonov}},
  \bibinfo{author}{\bibfnamefont{R.~D.} \bibnamefont{Coalson}},
  \bibnamefont{and}
  \bibinfo{author}{\bibfnamefont{Y.}~\bibnamefont{Dahnovsky}},
  \bibinfo{journal}{J. Chem. Phys.} \textbf{\bibinfo{volume}{116}},
  \bibinfo{pages}{10909} (\bibinfo{year}{2002}{\natexlab{a}}).

\bibitem[{\citenamefont{Tikhonov
  et~al.}(2002{\natexlab{b}})\citenamefont{Tikhonov, Coalson, and
  Dahnovsky}}]{tikhonov_2}
\bibinfo{author}{\bibfnamefont{A.}~\bibnamefont{Tikhonov}},
  \bibinfo{author}{\bibfnamefont{R.~D.} \bibnamefont{Coalson}},
  \bibnamefont{and}
  \bibinfo{author}{\bibfnamefont{Y.}~\bibnamefont{Dahnovsky}},
  \bibinfo{journal}{J. Chem. Phys.} \textbf{\bibinfo{volume}{117}},
  \bibinfo{pages}{567} (\bibinfo{year}{2002}{\natexlab{b}}).

\bibitem[{\citenamefont{Lehmann et~al.}(2003)\citenamefont{Lehmann, Kohler, and
  H\"anggi}}]{hanggi_1vs2}
\bibinfo{author}{\bibfnamefont{J.}~\bibnamefont{Lehmann}},
  \bibinfo{author}{\bibfnamefont{S.}~\bibnamefont{Kohler}}, \bibnamefont{and}
  \bibinfo{author}{\bibfnamefont{P.}~\bibnamefont{H\"anggi}},
  \bibinfo{journal}{J. Chem. Phys.} \textbf{\bibinfo{volume}{118}},
  \bibinfo{pages}{3283} (\bibinfo{year}{2003}).

\bibitem[{\citenamefont{Welack et~al.}(2006)\citenamefont{Welack, Schreiber,
  and Kleinekathofer}}]{welack}
\bibinfo{author}{\bibfnamefont{S.}~\bibnamefont{Welack}},
  \bibinfo{author}{\bibfnamefont{M.}~\bibnamefont{Schreiber}},
  \bibnamefont{and}
  \bibinfo{author}{\bibfnamefont{U.}~\bibnamefont{Kleinekathofer}},
  \bibinfo{journal}{J. Chem. Phys.} \textbf{\bibinfo{volume}{124}},
  \bibinfo{pages}{044712} (\bibinfo{year}{2006}).

\bibitem[{\citenamefont{Galperin and Nitzan}(2005)}]{galperin_05}
\bibinfo{author}{\bibfnamefont{M.}~\bibnamefont{Galperin}} \bibnamefont{and}
  \bibinfo{author}{\bibfnamefont{A.}~\bibnamefont{Nitzan}},
  \bibinfo{journal}{Phys. Rev. Lett.} \textbf{\bibinfo{volume}{95}},
  \bibinfo{pages}{206802} (\bibinfo{year}{2005}).

\bibitem[{\citenamefont{Galperin and Nitzan}(2006)}]{galperin_06}
\bibinfo{author}{\bibfnamefont{M.}~\bibnamefont{Galperin}} \bibnamefont{and}
  \bibinfo{author}{\bibfnamefont{A.}~\bibnamefont{Nitzan}},
  \bibinfo{journal}{J. Chem. Phys.} \textbf{\bibinfo{volume}{124}},
  \bibinfo{pages}{234709} (\bibinfo{year}{2006}).

\bibitem[{\citenamefont{Lehmann et~al.}(2004)\citenamefont{Lehmann, Kohler,
  May, and Hanggi}}]{lehmann_1vs2}
\bibinfo{author}{\bibfnamefont{J.}~\bibnamefont{Lehmann}},
  \bibinfo{author}{\bibfnamefont{S.}~\bibnamefont{Kohler}},
  \bibinfo{author}{\bibfnamefont{V.}~\bibnamefont{May}}, \bibnamefont{and}
  \bibinfo{author}{\bibfnamefont{P.}~\bibnamefont{Hanggi}},
  \bibinfo{journal}{J. Chem. Phys.} \textbf{\bibinfo{volume}{121}},
  \bibinfo{pages}{2278} (\bibinfo{year}{2004}).

\bibitem[{\citenamefont{Franco et~al.}(2007{\natexlab{b}})\citenamefont{Franco,
  Shapiro, and Brumer}}]{prlwire}
\bibinfo{author}{\bibfnamefont{I.}~\bibnamefont{Franco}},
  \bibinfo{author}{\bibfnamefont{M.}~\bibnamefont{Shapiro}}, \bibnamefont{and}
  \bibinfo{author}{\bibfnamefont{P.}~\bibnamefont{Brumer}},
  \bibinfo{journal}{Phys. Rev. Lett.} \textbf{\bibinfo{volume}{99}},
  \bibinfo{pages}{126802} (\bibinfo{year}{2007}{\natexlab{b}}).

\bibitem[{\citenamefont{Tully}(1998)}]{tully}
\bibinfo{author}{\bibfnamefont{J.~C.} \bibnamefont{Tully}}, in
  \emph{\bibinfo{booktitle}{Classical and Quantum Dynamics in Condensed Phase
  Simulations}}, edited by
  \bibinfo{editor}{\bibfnamefont{B.}~\bibnamefont{Berne}},
  \bibinfo{editor}{\bibfnamefont{G.}~\bibnamefont{Ciccotti}}, \bibnamefont{and}
  \bibinfo{editor}{\bibfnamefont{D.~F.} \bibnamefont{Coker}}
  (\bibinfo{publisher}{World Scientific}, \bibinfo{address}{Singapore},
  \bibinfo{year}{1998}), pp. \bibinfo{pages}{700--720}.

\bibitem[{\citenamefont{Halcomb and Diestler}(1986)}]{halcomb}
\bibinfo{author}{\bibfnamefont{L.~L.} \bibnamefont{Halcomb}} \bibnamefont{and}
  \bibinfo{author}{\bibfnamefont{D.~J.} \bibnamefont{Diestler}},
  \bibinfo{journal}{J. Chem. Phys.} \textbf{\bibinfo{volume}{84}},
  \bibinfo{pages}{3130} (\bibinfo{year}{1986}).

\bibitem[{\citenamefont{Bornemann et~al.}(1996)\citenamefont{Bornemann,
  Nettesheim, and Sch\"utte}}]{bornemann}
\bibinfo{author}{\bibfnamefont{F.~A.} \bibnamefont{Bornemann}},
  \bibinfo{author}{\bibfnamefont{P.}~\bibnamefont{Nettesheim}},
  \bibnamefont{and}
  \bibinfo{author}{\bibfnamefont{C.}~\bibnamefont{Sch\"utte}},
  \bibinfo{journal}{J. Chem. Phys.} \textbf{\bibinfo{volume}{105}},
  \bibinfo{pages}{1074} (\bibinfo{year}{1996}).

\bibitem[{\citenamefont{Prezhdo and Kisil}(1997)}]{prezdho}
\bibinfo{author}{\bibfnamefont{O.~V.} \bibnamefont{Prezhdo}} \bibnamefont{and}
  \bibinfo{author}{\bibfnamefont{V.~V.} \bibnamefont{Kisil}},
  \bibinfo{journal}{Phys. Rev. A} \textbf{\bibinfo{volume}{56}},
  \bibinfo{pages}{162} (\bibinfo{year}{1997}).

\bibitem[{\citenamefont{Fox}(2001)}]{fox}
\bibinfo{author}{\bibfnamefont{M.}~\bibnamefont{Fox}},
  \emph{\bibinfo{title}{Optical Properties of Solids}}
  (\bibinfo{publisher}{Oxford University Press}, \bibinfo{address}{New York},
  \bibinfo{year}{2001}).

\bibitem[{\citenamefont{Hellums and Frensley}(1994)}]{frensley}
\bibinfo{author}{\bibfnamefont{J.~R.} \bibnamefont{Hellums}} \bibnamefont{and}
  \bibinfo{author}{\bibfnamefont{W.~R.} \bibnamefont{Frensley}},
  \bibinfo{journal}{Phys. Rev. B} \textbf{\bibinfo{volume}{49}},
  \bibinfo{pages}{2904} (\bibinfo{year}{1994}).

\bibitem[{\citenamefont{Negele and Orland}(1998)}]{negele}
\bibinfo{author}{\bibfnamefont{J.~W.} \bibnamefont{Negele}} \bibnamefont{and}
  \bibinfo{author}{\bibfnamefont{H.}~\bibnamefont{Orland}},
  \emph{\bibinfo{title}{Quantum Many-Particle Systems}}
  (\bibinfo{publisher}{Westview Press}, \bibinfo{address}{Boulder, CO},
  \bibinfo{year}{1998}).

\bibitem[{\citenamefont{Abramowitz and Stegun}(1965)}]{bessel:abramowitz}
\bibinfo{editor}{\bibfnamefont{M.}~\bibnamefont{Abramowitz}} \bibnamefont{and}
  \bibinfo{editor}{\bibfnamefont{I.~A.} \bibnamefont{Stegun}}, eds.,
  \emph{\bibinfo{title}{Handbook of Mathematical Functions}}
  (\bibinfo{publisher}{Dover}, \bibinfo{address}{New York},
  \bibinfo{year}{1965}), chap. \bibinfo{chapter}{9.1.21}.

\bibitem[{\citenamefont{Brankin et~al.}(1992)\citenamefont{Brankin, Gladwell,
  and Shampine}}]{rksuite}
\bibinfo{author}{\bibfnamefont{R.~W.} \bibnamefont{Brankin}},
  \bibinfo{author}{\bibfnamefont{I.}~\bibnamefont{Gladwell}}, \bibnamefont{and}
  \bibinfo{author}{\bibfnamefont{L.~F.} \bibnamefont{Shampine}},
  \emph{\bibinfo{title}{RKSUITE: a suite of Runge-Kutta codes for the initial
  value problem for ODEs}} (\bibinfo{publisher}{Softreport 92-S1},
  \bibinfo{address}{Department of Mathematics, Southern Methodist University,
  Dallas, Texas}, \bibinfo{year}{1992}), \bibinfo{note}{www.netlib.org}.

\bibitem[{\citenamefont{Wu et~al.}(2006)\citenamefont{Wu, Ogawa, and
  Ho}}]{wusw}
\bibinfo{author}{\bibfnamefont{S.~W.} \bibnamefont{Wu}},
  \bibinfo{author}{\bibfnamefont{N.}~\bibnamefont{Ogawa}}, \bibnamefont{and}
  \bibinfo{author}{\bibfnamefont{W.}~\bibnamefont{Ho}},
  \bibinfo{journal}{Science} \textbf{\bibinfo{volume}{312}},
  \bibinfo{pages}{1362} (\bibinfo{year}{2006}).

\bibitem[{\citenamefont{Chen et~al.}(2006)\citenamefont{Chen, Shakya, and
  Lipson}}]{lipsonmichal}
\bibinfo{author}{\bibfnamefont{L.}~\bibnamefont{Chen}},
  \bibinfo{author}{\bibfnamefont{J.}~\bibnamefont{Shakya}}, \bibnamefont{and}
  \bibinfo{author}{\bibfnamefont{M.}~\bibnamefont{Lipson}},
  \bibinfo{journal}{Opt. Lett.} \textbf{\bibinfo{volume}{31}},
  \bibinfo{pages}{2133} (\bibinfo{year}{2006}).

\bibitem[{\citenamefont{Aeschlimann et~al.}(2007)\citenamefont{Aeschlimann,
  Bauer, Bayer, Brixner, Garcia~de Abajo, Pfeiffer, Rohmer, Spindler, and
  Steeb}}]{aeschlimann}
\bibinfo{author}{\bibfnamefont{M.}~\bibnamefont{Aeschlimann}},
  \bibinfo{author}{\bibfnamefont{M.}~\bibnamefont{Bauer}},
  \bibinfo{author}{\bibfnamefont{D.}~\bibnamefont{Bayer}},
  \bibinfo{author}{\bibfnamefont{T.}~\bibnamefont{Brixner}},
  \bibinfo{author}{\bibfnamefont{F.~J.} \bibnamefont{Garcia~de Abajo}},
  \bibinfo{author}{\bibfnamefont{W.}~\bibnamefont{Pfeiffer}},
  \bibinfo{author}{\bibfnamefont{M.}~\bibnamefont{Rohmer}},
  \bibinfo{author}{\bibfnamefont{C.}~\bibnamefont{Spindler}}, \bibnamefont{and}
  \bibinfo{author}{\bibfnamefont{F.}~\bibnamefont{Steeb}},
  \bibinfo{journal}{Nature} \textbf{\bibinfo{volume}{446}},
  \bibinfo{pages}{301} (\bibinfo{year}{2007}).

\bibitem[{\citenamefont{Sussman et~al.}(2006)\citenamefont{Sussman, Townsend,
  Ivanov, and Stolow}}]{sussman}
\bibinfo{author}{\bibfnamefont{B.~J.} \bibnamefont{Sussman}},
  \bibinfo{author}{\bibfnamefont{D.}~\bibnamefont{Townsend}},
  \bibinfo{author}{\bibfnamefont{M.~Y.} \bibnamefont{Ivanov}},
  \bibnamefont{and} \bibinfo{author}{\bibfnamefont{A.}~\bibnamefont{Stolow}},
  \bibinfo{journal}{Science} \textbf{\bibinfo{volume}{314}},
  \bibinfo{pages}{278} (\bibinfo{year}{2006}).

\end{thebibliography}

\end{document}